\documentclass[pdfa,a4paper,UKenglish,cleveref,autoref,thm-restate]{lipics-v2021}
\nolinenumbers

\usepackage[ruled,vlined,linesnumbered]{algorithm2e}
\usepackage{graphicx,booktabs,amsmath}
\newcommand{\bm}[1]{\mathbf{#1}}

\renewcommand{\bm}[1]{\mathbf{#1 }}
\newcommand{\allattr}{{\mathbb A}}
\newcommand{\allrel}{{\mathbb R}}
\newcommand{\attr}{{\texttt{attr}}}
\newcommand{\head}{{\texttt{head}}}
\newcommand{\rel}{{\texttt{rels}}}
\newcommand{\dom}{{\texttt{dom}}}
\newcommand{\DP}{\texttt{DP}}
\newcommand{\ADP}{\texttt{ADP}}
\newcommand{\ourprob}{{\texttt{SWP}}}
\newcommand{\cover}{{Q_\textsf{cover}}}
\renewcommand{\matrix}{{Q_\textsf{matrix}}}
\newcommand{\pyramid}{{Q_\textsf{pyramid}}}
\renewcommand{\mod}{\textsf{mod }}
\renewcommand{\line}{{Q_\textsf{line3}}}
\newcommand{\rev}[1]{{\color{black}{#1}}}

\theoremstyle{definition}
\newtheorem{remarknew}{Remark}

\title{Finding Smallest Witnesses for Conjunctive Queries}

\author{Xiao Hu}{University of Waterloo, Canada}{xiaohu@uwaterloo.ca}{https://orcid.org/0000-0002-7890-665X}{}
\author{Stavros Sintos}{University of Illinois Chicago, IL, USA}{stavros@uic.edu}{https://orcid.org/0000-0002-2114-8886}{}

\authorrunning{X. Hu and S. Sintos}
\Copyright{Xiao Hu and Stavros Sintos}



\relatedversiondetails{Full Version}{https://arxiv.org/abs/2311.18157}

\EventEditors{Graham Cormode and Michael Shekelyan}
\EventNoEds{2}
\EventLongTitle{27th International Conference on Database Theory (ICDT 2024)}
\EventShortTitle{ICDT 2024}
\EventAcronym{ICDT}
\EventYear{2024}
\EventDate{March 25--28, 2024}
\EventLocation{Paestum, Italy}
\EventLogo{}
\SeriesVolume{290}
\ArticleNo{24}

\begin{document}

\maketitle
\vspace{0.5\baselineskip}
\enlargethispage{-0.5\baselineskip}
\begin{abstract}
A witness is a sub-database that preserves the query results of the original database but of much smaller size. It has wide applications in query rewriting and debugging, query explanation, IoT analytics, multi-layer network routing, etc. In this paper, we study the smallest witness problem (\ourprob) for the class of conjunctive queries (CQs) without self-joins. 

We first establish the dichotomy that \ourprob\ for a CQ can be computed in polynomial time if and only if it has {\em head-cluster property}, unless $\texttt{P} = \texttt{NP}$. We next turn to the approximated version by relaxing the size of a witness from being minimum. We surprisingly find that the {\em head-domination} property -- that has been identified for the deletion propagation problem \cite{kimelfeld2012maximizing} -- can also precisely capture the hardness of the approximated smallest witness problem. In polynomial time, \ourprob\ for any CQ with head-domination property can be approximated within a constant factor, while \ourprob\ for any CQ without such a property cannot be approximated within a logarithmic factor, unless $\texttt{P} = \texttt{NP}$. 

We further explore efficient approximation algorithms for CQs without head-domination property: (1) we show a trivial algorithm which achieves a polynomially large approximation ratio for general CQs; (2) for any CQ with only one non-output attribute, such as star CQs, we show a greedy algorithm with a logarithmic approximation ratio; (3) for line CQs, which contain at least two non-output attributes, we relate \ourprob\ problem to the directed steiner forest problem, whose algorithms can be applied to line CQs directly. Meanwhile, we establish a much higher lower bound, exponentially larger than the logarithmic lower bound obtained above. It remains open to close the gap between the lower and upper bound of the approximated \ourprob\ for CQs without head-domination property.
\end{abstract}

\section{Introduction}
\label{sec:intro}

To deal with large-scale data in analytical applications, people have developed a large body of data summarization techniques to reduce computational as well as storage complexity, such as sampling~\cite{chaudhuri1999random, zhao2018random, chen2020random}, sketch~\cite{cormode2020small}, coreset~\cite{phillips2017coresets} and factorization~\cite{olteanu2016factorized}.  \rev{The notion of witness has been studied as one form of why-provenance~\cite{buneman2001and, green2007provenance, amsterdamer2011provenance} that provides a {\em proof} for output results, with wide applications in explainable data-intensive analytics.} The {\em smallest witness problem} was first proposed by~\cite{miao2019explaining} \rev{that given a query $Q$, a database $D$ and one specific query result $t \in Q(D)$, the target is to find the smallest sub-database $D' \subseteq D$ such that $t$ is witnessed by $D'$, i.e., $t \in Q(D')$.} In this paper, we consider a generalized notion for all query results, i.e., our target is to find the smallest sub-database $D' \subseteq D$ such that all query results can be witnessed by $D'$, i.e., $Q(D) = Q(D')$. \rev{Our generalized smallest witness has many useful applications in practice, such as helping students learn SQL queries~\cite{miao2019explaining}, query rewriting, query explanation, multi-layer network routing, IoT analytics on edge devices\cite{paparrizos2021vergedb}.} We mention three application scenarios: 
\begin{example}
\label{ex:1}
Alice located at Seattle wants to \rev{send} the query results of $Q$ over a database $D$ (which is also stored at Seattle) to Bob located at New York. Unfortunately, the \rev{number of} query results could be \rev{polynomially} large in terms of the number of tuples in $D$. An alternative is to send the entire database $D$, since Bob can retrieve all query results by executing $Q$ over $D$ at New York. However, moving the entire database is also expensive. A natural question \rev{arises}: \rev{is it necessary for Alice to send the entire $D$ or $Q(D)$}? If not, what is the smallest subset of tuples to send? 
\end{example}

\begin{example}
\label{ex:2}
\rev{Charlie is a novice at learning SQL in a undergraduate database course. Suppose there is a huge test database $D$, a correct query $Q$ that Charlie is expected to learn, and a wrong query $Q'$ submitted by him, where some answers in $Q(D)$ are missed from $Q'(D)$. To help Charlie debug, the instructor can simply show the whole test database $D$ to him. However, Charlie will have to dive into such a huge database to figure out where his query goes wrong. A natural question arises: is it necessary to show the entire $D$ to Charlie? If not, what is the smallest subset of tuples to show so that Charlie can quickly find all missing answers by his wrong query?}
\end{example}

\begin{example}
\label{ex:3}
\rev{
In a multi-layer communication network, clients and servers are connected by routers organized into layers, such that links (or edges) exist between routers residing in consecutive layers. What is the smallest subset of links needed for building a fully connected network, i.e., every client-server pair is connected via a directed path? For a given network, what is the maximum number of links that can be broken while the connectivity with respect to the client-server pairs does not change? This information could help evaluate the inherent robustness of a network to either malicious attacks or even just random failures.
}
\end{example}

Recall that our smallest witness problem finds the smallest sub-database $D' \subseteq D$ such that $Q(D) = Q(D')$. \rev{It would be sufficient for Alice to send $D'$, while Bob can retrieve all query results by executing $Q$ over $D'$, saving much transmission cost. Also, it would be sufficient for the instructor to show $D'$ to Charlie, from which all correct answers in $Q(D)$ can be recovered, saving Charlie much efforts in exploring a huge test database.\footnote{\rev{The smallest witness for a single query result \cite{miao2019explaining} has been incorporated into an educational tool (\url{https://dukedb-hnrq.github.io/}), successfully employed in Duke database courses with 1,000+ undergraduate users. Our generalized version can also be incorporated and save more efforts by showing one small witness for all answers.}} Moreover, the connectivity of a multi-layer network $D$ can be modeled as a line query $Q$ (formally defined in Section~\ref{sec:line}) with connected client-server pairs as $Q(D)$, such that $D'$ is a smallest subset of links needed for maintaining the desired connectivity, and $D - D'$ is a maximum subset of links that can be removed safely.}
In this paper, we aim to design algorithms that can efficiently compute or approximate the smallest witness for conjunctive queries, and understand the hardness of this problem when such algorithms do not exist. 

\subsection{Problem Definition}
\label{sec:definition}
\enlargethispage{\baselineskip}
Let $\allrel$ be a database schema that contains $m$ relations $R_1, R_2, \cdots, R_m$. Let $\allattr$ be the set of all attributes in $\allrel$. Each relation $R_i$ is defined on a subset of attributes $\allattr_i \subseteq \allattr$. We use $A, B, C, A_1, A_2, A_3, \cdots$ etc. to denote the attributes in $\allattr$ and $a, b, c, \cdots$ etc. to denote their values. Let $\dom(A)$ be the domain of attribute \rev{$A \in \allattr$}. The domain of a set of attributes $X \subseteq \allattr$ is defined as $\dom(X) = \prod_{A \in X} \dom(A)$. Given the database schema $\allrel$, let $D$ be a given database of $\allrel$, and let the corresponding \rev{relations} of $R_1, \cdots, R_m$ be $R_1^{D}, \cdots$, $R_m^{D}$, where $R_i^D$ is a collection of tuples defined on $\dom(\allattr_i)$. The {\em input size} of database $D$ is denoted as $N = |D|= \sum_{i \in [m]}|R^D_i|$. Where $D$ is clear from the context, we will drop the superscript. \pagebreak

\begin{figure}[h]
\centering
\begin{tabular}{|c|c|}
\multicolumn{2}{c}{$R_1$}\\
\hline
A & B \\\hline\hline
a1 & b1 \\
a2 & b2 \\
a3 & b2 \\\hline
\end{tabular}
\begin{tabular}{|c|c|}
\multicolumn{2}{c}{$R_2$}\\
\hline
B & C \\\hline\hline
b1 & c1 \\
b2 & c3 \\
b3 & c2 \\
b3 & c3 \\
\hline
\end{tabular}
\begin{tabular}{|c|c|}
\multicolumn{2}{c}{$R_3$}\\
\hline
C & F \\\hline\hline
c1 & f1 \\
c2 & f3 \\
c3 & f3 \\
\hline
\end{tabular}
\begin{tabular}{|c|c|}
\multicolumn{2}{c}{$R_4$}\\
\hline
C & H \\\hline\hline
c1 & h1 \\
c2 & h1 \\
c3 & h1 \\
c3 & h2 \\
\hline
\end{tabular}
\begin{tabular}{|c|c|c|}
\multicolumn{3}{c}{$Q(D)$}\\
\hline
A & C & F\\\hline\hline
a1 & c1  & f1\\
a2 & c3  & f3\\
a3 & c3  & f3\\
\hline
\end{tabular}
\caption{An example of database schema $\allrel = \{R_1,R_2,R_3,R_4\}$ (with $\allattr$ $= \{A, B, C, F, H\}$, $\attr(R_1) =\{A, B\}$, $\attr(R_2)=\{B, C\}$, $\attr(R_3) =\{C, F\}$ and $\attr(R_4) = \{C,H\}$), a database $D$, and the result of CQ $Q(A,C,F):-R_1(A,B), R_2(B,C), R_3(C,F), R_4(C,H)$ over $D$. $D' = \{(a_1, b_1), (b_1,c_1), (c_1, f_1),(c_1, h_1)\}$ is the solution to $\ourprob(Q,D, \langle a_1, c_1, f_1 \rangle)$. $D'$ together with tuples $\{(a_2, b_2), (a_3, b_2), (b_2, c_3), (c_3, f_3), (c_3, h_2)\}$ is the solution to $\ourprob(Q,D)$. }  
\label{fig:example_setup}
\end{figure}

We consider the class of \emph{conjunctive queries without self-joins}:  \[Q(\bm{A}): -R_1(\mathbb{A}_1), R_2(\mathbb{A}_2), \cdots, R_m(\mathbb{A}_m)\] where $\bm{A} \subseteq \mathbb{A}$ is the set of {\em output attributes} (a.k.a. {\em free attributes}) and $\mathbb{A} - \bm{A}$ is the set of {\em non-output attributes} (a.k.a. {\em existential attributes}). A CQ is {\em full} if $\bm{A} = \mathbb{A}$, indicating the natural join among the given relations; otherwise, it is {\em non-full}. Each $R_i$ in $Q$ is distinct. When a CQ $Q$ is evaluated on database $D$, its query result denoted as $Q(D)$ is the projection of natural join result of $R_1(\mathbb{A}_1)\Join R_2(\mathbb{A}_2) \Join \cdots \Join R_m(\mathbb{A}_m)$ onto $\bm{A}$ (after removing duplicates). 

\begin{definition}[{\sc Smallest Witness Problem (\ourprob)}]
\label{def:SWP}		
For CQ $Q$ and database $D$, it asks to find a subset of tuples $D' \subseteq D$ such that $Q(D) = Q(D')$, while there exists no subset of tuples $D'' \subseteq D$ such that $Q(D'') = Q(D)$ and $|D''| < |D'|$.	
\end{definition}
Given $Q$ and $D$, we denote the above problem by $\ourprob(Q, D)$. See an example in Figure~\ref{fig:example_setup}. \rev{We note that the solution to $\ourprob(Q, D)$ may not be unique, hence our target simply finds one such solution.} We study the data complexity~\cite{vardi1982complexity} of \ourprob\, i.e., the sizes of database schema and query are considered as constants, and the complexity is in terms of input size $N$. For any CQ $Q$ and database $D$, the size of query results  $|Q(D)|$ is polynomially large in terms of $N$, and $Q(D)$ can also be computed in polynomial time in terms of $N$. In contrast, the size of \ourprob$(Q,D)$ is always smaller than $N$, while as we see later \ourprob$(Q,D)$ may not be computed in polynomial time in terms of $N$. Again, our target is to compute the smallest witness instead of the query results. We say that \ourprob\ is {\em poly-time solvable} for $Q$ if, \rev{for an arbitrary database $D$}, $\ourprob(Q,D)$ can be computed in polynomial time in terms of $|D|$. As shown later, $\ourprob$ is not poly-time solvable for a large class of CQs, so we introduce an approximated version:
\begin{definition}[{\sc $\theta$-Approximated Smallest Witness Problem (\texttt{ASWP})}]
\label{def:ASWP}		
For CQ $Q$ and database $D$, it asks to find a subset of tuples $D' \subseteq D$ such that $Q(D') = Q(D)$ and $|D'| \le \theta \cdot |D^*|$, where $D^*$ is \rev{a} solution to \ourprob$(Q,D)$.
\end{definition}

{Also, \ourprob\ is $\theta$-approximable for $Q$ if, for an arbitrary \rev{database $D$}, there is a $\theta$-approximated solution to $\ourprob(Q,D)$ that can be computed in polynomial time in terms of $|D|$.}

\subsection{Our Results}
\label{sec:contribution}
\begin{figure}[t]
\centering
\includegraphics[scale=0.85]{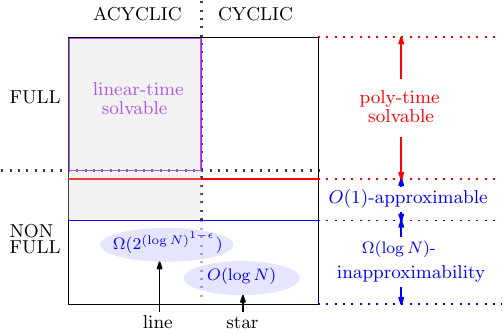}
\caption{Summary of our results. The shadow area is the class of free-connex CQs.  }
\label{fig:summary}
\vspace{-1em}
\end{figure}

Our main results obtained can be summarized as (see Figure~\ref{fig:summary}):

\enlargethispage{\baselineskip}
In Section~\ref{sec:exact}, we obtain a dichotomy of computing \ourprob\ for CQs. More specifically, \ourprob\ for any CQ with head-cluster property (Definition~\ref{def:head-cluster}) can be solved by a trivial poly-time algorithm, while \ourprob\ for any CQ without head-cluster property is NP-hard by resorting to the {\em NP-hardness of set cover} problem (Section~\ref{sec:exact-hardness}). 

In Section~\ref{sec:approximate}, we show a dichotomy of approximating \ourprob\ for CQs without head-cluster property. The {\em head-domination} property that has been identified for {\em deletion propagation} problem~\cite{kimelfeld2012maximizing}, also captures the hardness of approximating \ourprob. We show a poly-time algorithm that can return a $O(1)$-approximated solution to $\ourprob$ for CQs with head-domination property. On the other hand, we prove that  \ourprob\ cannot be approximated within a factor of $(1-o(1)) \cdot \log N$ for every CQ without head-domination property, unless \texttt{P} = \texttt{NP}, by resorting to the {\em logarithmic inapproximability of set cover} problem (in Section~\ref{sec:approximate-hard}). Interestingly, this separation on approximating \ourprob\ for {\em acyclic} CQs (in Section~\ref{sec:classification}) coincides with the separation of {\em free-connex} and {\em non-free-connex} CQs  (in Section~\ref{sec:approximate-easy}) in the literature.
In Section~\ref{sec:approximation-algorithm}, we further explore approximation algorithms for CQs without head-domination property. Firstly, we show a baseline of returning the union of witnesses for every query result leads to a $O(N^{1-1/\rho^*})$-approximated solution, where $\rho^*$ is the fractional edge covering number\footnote{For a CQ $Q$, a fractional edge covering is a function $W: \rel(Q) \to [0,1]$ with $\sum_{R_i: A \in \attr(R_i)} W(R_i) \ge 1$ for every $A \in \allattr$. The fractional edge covering number is the minimum value of $\sum_{R_i: R_i \in \rel(Q)} W(R_i)$ over all fractional edge coverings. } of the input CQ~\cite{AGM2008}. Furthermore, for any CQ with only one non-output attribute, which includes the commonly-studied {\em star} CQs, we show a greedy algorithm that can approximate \ourprob\ within a $O(\log N)$ factor, matching the lower bound. However, for another commonly-studied class of {\em line} CQs, which contains more than two non-output attributes, we prove a much higher lower bound $\Omega(2^{(\log N)^{1-\epsilon}})$ for any $\epsilon > 0$ in approximating \ourprob, by resorting to the {\em label cover} problem (see Appendix~\ref{appendix:approximation-algorithm}). Meanwhile, we observe that \ourprob\ problem for line queries is a special case of the {\em directed Steiner forest} (DSF) problem (Section~\ref{sec:line}), and therefore existing algorithms for DSF can be applied to \ourprob\ directly. But, how to close the gap between the upper and lower bounds on approximating \ourprob\ for line CQs remains open.  {}
\section{Preliminaries}
\label{sec:prelim}
\subsection{Notations and Classifications of CQs} \label{sec:classification}
\enlargethispage{\baselineskip}
Extending the notation in Section~\ref{sec:definition}, we use $\rel(Q)$ to denote all the relations that appear in the body of $Q$, and use $\attr(Q), \head(Q) \subseteq \attr(Q)$ to denote all the attributes 
that appear in the body, head of $Q$ separately (so, $\head(Q) = \bm{A}$ in Section~\ref{sec:definition}). Moreover, $\head(R_i)=\head(Q) \cap \attr(R_i)$. For any attribute $A \in \attr(R_i)$, $ \pi_{A} t$ denotes the value over attribute $A$ of tuple $t$. Similarly, for a set of attributes $X \subseteq \attr(R_i)$, $\pi_X t$ denotes values over attributes in $X$ of tuple $t$. We also mention two important classes of CQs. 
\begin{definition}[Acyclic CQs~\cite{beeri1983desirability, fagin1983degrees}]
A CQ $Q$ is acyclic if there exists a tree $\mathbb{T}$ such that (1) there is a one-to-one correspondence between the nodes of $\mathbb{T}$ and relations in $Q$; and (2) for every attribute $A \in \attr(Q)$, the set of nodes containing $A$ forms a connected subtree of $\mathbb{T}$. Such a tree is called the {\em join tree} of $Q$.
\end{definition} 

\begin{definition}[Free-connex CQs~\cite{bagan2007acyclic}]
A CQ $Q$ is free-connex if $Q$ is acyclic and the resulted CQ by adding another relation contains exactly $\head(Q)$ to $Q$ is also acyclic.  
\end{definition}

\subsection{\ourprob\ for One Query Result}
The \ourprob\ problem for one query result is formally defined as:
\begin{definition}[{\sc \ourprob\ for One Query Result}]
\label{def:swp}		
For CQ $Q$, database $D$ and query result $t \in Q(D)$, it asks for finding a subset of tuples $D' \subseteq D$ such that $t \in Q(D')$, while there is no subset $D'' \subseteq D$ such that $t \in Q(D'')$ and $|D''| < |D'|$.
\end{definition}

Given $Q$, $D$ and $t$, we denote the above problem by $\ourprob(Q, D, t)$. It has been shown by~\cite{miao2019explaining} that $\ourprob(Q,D,t)$ can be computed in polynomial time for arbitrary $Q$, $D$, and $t\in Q(D)$. Their algorithm~\cite{miao2019explaining} simply finds an arbitrary full join result $t' \in \Join_{R_i \in \rel(Q)} R_i$ such that $\pi_{\head(Q)} t' = t$, and returns all participating tuples in $\left\{\pi_{\attr(R_i)} t': R_i \in \rel(Q)\right\}$ as the smallest witness for $t$. This primitive is used in building our \ourprob\ algorithm. The \ourprob\ problem for a Boolean CQ ($\bm{A} = \emptyset$, indicating whether the result of underlying natural join is empty or not) can be solved by finding \ourprob\ for an arbitrary join result in its full version. 

\subsection{Notions of Connectivity}
\begin{figure}[t]
\centering
\vspace{1em}
\includegraphics[scale=0.9]{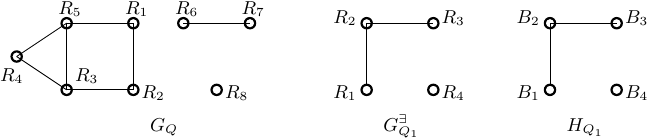}
\vspace{-1em}
\caption{An illustration of $G_Q$ for $Q(A_1,A_2,A_3, A_4,A_5):-$ $R_1(A_1,B_1)$, $R_2(B_1,B_2)$, $R_3(A_2,B_2,B_3)$, $R_4(A_2,A_3,B_4)$, $R_5(A_1,A_2)$, $R_6(A_4, B_5)$, $R_7(B_5,A_5)$, $R_8(B_6,B_7)$ with three subqueries $Q_1(A_1,A_2,A_3):-$ $R_1(A_1,B_1)$, $R_2(B_1,B_2)$, $R_3(A_2,B_2,B_3)$, $R_4(A_2,A_3,B_4)$, $R_5(A_1,A_2)$, and $Q_2(A_4,A_5):-$ $R_6(A_4, B_5)$, $R_7(B_5,A_5)$ and $Q_3:-R_8(B_6,B_7)$. The middle is $G^\exists_{Q_1}$ for $Q_1$, with two connected components $\{R_1,R_2,R_3\}$, $\{R_4\}$ and dominants $R_5$, $R_4$. The right is $H_{Q_1}$ for $Q_1$, with two connected components $\{B_1,B_2,B_3\}, \{B_4\}$.}
\label{fig:connectivity}
\end{figure}

We give three important notions of connectivity for CQs, which will play an important role in characterizing the structural properties used in \ourprob. See an example in Figure~\ref{fig:connectivity}.

\subparagraph{Connectivity of CQ.} We capture the {\em connectivity} of a CQ $Q$ by modeling it as a graph $G_Q$, where each relation $R_i$ is a vertex and there is an edge between $R_i, R_j \in \rel(Q)$ if $\attr(R_i) \cap \attr(R_j) \neq \emptyset$. A CQ $Q$ is {\em connected} if $G_Q$ is connected, and {\em disconnected} otherwise. For a disconnected CQ $Q$, we can decompose it into multiple connected subqueries by applying search algorithms
on $G_Q$, and finding all connected components for $G_Q$. The
set of relations corresponding to the set of vertices in one connected
component of $G_Q$ form a connected subquery of $Q$.  Given a disconnected CQ $Q$, let $Q_1,Q_2,\cdots,Q_k$ be its connected subqueries. Given a database $D$ over $Q$, let $D_i\subseteq D$ be the corresponding sub-databases defined for $Q_i$. Observe that every witness for $Q(D)$ is the disjoint union of a witness for $Q_i(D_i)$, for $i \in [k]$. Hence, Lemma~\ref{lem:disconnected} follows. In the remaining of this paper, we assume that $Q$ is connected.

\begin{lemma}
\label{lem:disconnected}
For a disconnected CQ $Q$ of $k$ connected components $Q_1, Q_2, \cdots, Q_k$, $\ourprob$ is poly-time solvable for $Q$ if and only if $\ourprob$ is poly-time solvable for \rev{every $Q_i$}, where $i \in [k]$.
\end{lemma}

\subparagraph{Existential-Connectivity of CQ.} We capture the existential-connectivity of a CQ $Q$ by modeling it as a graph $G^\exists_Q$, where each relation $R_i$ with $\attr(R_i) - \head(Q) \neq \emptyset$ is a vertex, and there is an edge between $R_i, R_j \in \rel(Q)$ if $\attr(R_i) \cap \attr(R_j) - \head(Q) \neq \emptyset$. We can find the connected components of $G^\exists_Q$ by applying search algorithm on $G^\exists_Q$, and finding all connected components for $G^\exists_Q$. Let $E_1,E_2,\cdots, E_k \subseteq \rel(Q)$ be the connected components of $G^\exists_Q$, each corresponding to a subset of relations in $Q$. 

\subparagraph{Nonout-Connectivity of CQ.} We capture the nonout-connectivity of a CQ $Q$ by modeling it as a graph $H_Q$, where each non-output attribute $A \in \attr(Q)-\head(Q)$ is a vertex, and there is an edge between $A,B \in \attr(Q) -\head(Q)$ if there exists a relation $R_i \in \rel(Q)$ such that $A,B \in \attr(R_i)$.  We can find the connected components of $H_Q$, and finding all connected components for $H_Q$. Let $H_1, H_2, \cdots, H_k \subseteq \attr(Q) - \head(Q)$ be the connected components of $H_Q$, each corresponding to a subset of non-output attributes in $Q$.

\subsection{Head Cluster and Domination} 

These two important structural properties identified for characterizing the hardness of (\texttt{A})\ourprob\ are directly built on the existential-connectivity of a CQ $Q$ and the notion of dominant relation. For a CQ $Q$ with a subset $E \subseteq \rel(Q)$ of relations, $R_i \in \rel(Q)$ is a {\em dominant} relation for $E$ if every output attribute appearing in any relation of $E$ also appears in $R_i$, i.e., $\cup_{R_j \in E} \head(R_j) \subseteq \head(R_i)$. 

\begin{definition}[Head Domination~\cite{KimelfeldVW11}]
\label{def:head-domination}
For CQ $Q$, let $E_1, E_2,\cdots, E_k$ be the connected components of $G^\exists_Q$. $Q$ has head-domination property if for any $i \in [k]$, there exists a dominant relation from $\rel(Q)$ for $E_i$. \end{definition}

The notion of head-domination property has been first identified for deletion propagation with side effect problem~\cite{kimelfeld2012maximizing}, which studied the smallest number of tuples to remove so that a subset of desired query results must disappear while maintain as many as remaining query results. We give a detailed comparison between \ourprob\ and deletion propagation in Appendix~\ref{appendix:deletion-propagation}, although they solve completely independent problem for CQs without self-joins.

\begin{definition}[Head Cluster]
\label{def:head-cluster}
For CQ $Q$, let $E_1, \cdots, E_k$ be connected components of $G^\exists_Q$. $Q$ has head-cluster property if for any $i \in [k]$, every $R_j \in E_i$ is a dominant relation for $E_i$. \end{definition}

There is an equivalent but simpler definition for head-cluster property: A CQ $Q$ has head-cluster property if for every pair of relations $R_i, R_j \in \rel(Q)$ with \rev{$\head(R_i) \neq \head(R_j)$}, there must be $\attr(R_i) \cap \attr(R_j) \subseteq \head(Q)$. Here, we define head-cluster property based on dominant relation, since it is a special case of head-domination property. 

\section{Dichotomy of Exact \ourprob}
\label{sec:exact}

In this section, we focus on computing \ourprob\ exactly for CQs, which can be efficiently done if {\em head-cluster} property is satisfied. \rev{All missing proofs are given in Appendix~\ref{appendix:SWP-easy}.} 

\begin{theorem}\label{the:dichotomy-exact}
If a CQ $Q$ has head-cluster property, $\ourprob$ is poly-time solvable; otherwise, $\ourprob$ is not poly-time solvable, unless $\texttt{P} = \texttt{NP}$.
\end{theorem}

\subsection{An Exact Algorithm}
\label{sec:algo}
\begin{algorithm}[t]
\caption{{\sc \ourprob}$(Q,D)$.}
\label{alg:easy-SWP}
$D' \gets \emptyset$\;
$(E_1, E_2, \cdots, E_k) \gets$ connected components of $G^\exists_Q$\; 
${\bf A}_1, {\bf A}_2, \cdots, {\bf A}_k \gets $ output attributes of $E_1, E_2, \cdots, E_k$\;
\ForEach{$R_j \in \rel(Q)$ with $\attr(R_j) \subseteq \head(Q)$}{
$D' \gets D' \cup \pi_{\attr(R_j)} Q(D)$\;
}
\ForEach{$i \in [k]$}{
Define $Q_i\left({\bf A}_i\right):- \{R_j(\allattr_{j}): R_j \in E_i\}$\;
\ForEach{$t' \in \pi_{{\bf A}_i}Q(D)$}{
$D' \gets D' \biguplus \ourprob\left(Q_i, \{R_j: R_j \in E_i\}, t'\right)$\;
}
}
\Return $D'$\;
\end{algorithm}

We prove the first part of Theorem~\ref{the:dichotomy-exact} with a poly-time algorithm. The head-cluster property implies that if two relations have different output attributes, they share no common non-output attributes. This way, we can cluster relations by output attributes.
As shown, Algorithm~\ref{alg:easy-SWP} partitions all relations into $\{E_1, E_2, \cdots, E_k\}$ based on the connected components in $G^\exists_Q$. For the subset of relations in one connected component $E_i$, every relation is a dominant relation, i.e., shares the same output attributes. If one relation only contains output attributes $\bm{A}_i$ (line 4), it must appear alone as a singleton component, since we assume there is no duplicate relations in the input CQ. All tuples from such a relation that participate in any query results must be included by every witness to $Q(D)$. We next consider the remaining components containing at least two relations. 
\rev{In $E_i$, for each tuple $t' \in \pi_{{\bf A}_i} Q(D)$ in the projection of query results onto the output attributes $\bm{A}_i$, Algorithm~\ref{alg:easy-SWP} computes the smallest witness for $t'$ in sub-query $Q_i$ defined on relations in $E_i$. The disjoint union ($\biguplus$) of witnesses returned for all groups forms the final witness.} On a CQ with head-cluster property, Algorithm~\ref{alg:easy-SWP} can be stated in a simpler way (see Appendix~\ref{appendix:SWP-easy}). Algorithm~\ref{alg:easy-SWP} runs in polynomial time, as (\romannumeral 1) $Q(D)$ can be computed in polynomial time; (\romannumeral 2) $|Q(D)|$ is polynomially large; and (\romannumeral 3) the primitive in line 9 only takes $O(1)$ time. 

\rev{\begin{lemma}\label{lem:easy}
For a CQ $Q$ with head-cluster property, Algorithm~\ref{alg:easy-SWP} finds a solution to \ourprob$(Q,D)$ for any database $D$ in polynomial time.
\end{lemma}}
\vspace{-0.9\baselineskip}
\begin{proof}
\enlargethispage{1.7\baselineskip}
We prove that for any CQ $Q$ with head-cluster property and an arbitrary database $D$, Algorithm~\ref{alg:easy-SWP} returns the solution to $\ourprob(Q,D)$. Together with the fact that Algorithm~\ref{alg:easy-SWP} runs in polynomial time, we finish the proof for Lemma~\ref{lem:easy}. Let $D'$ be the solution returned by Algorithm~\ref{alg:easy-SWP}. Let $D'_i \subseteq D'$ be the set of tuples from relations in $E_i$. We show that $D'$ is a witness for $Q(D)$, i.e., $Q(D')=Q(D)$.

%\underline{\em Direction $\subseteq$.}
\vspace{-0.9\baselineskip}
\proofsubparagraph*{\boldmath Direction $\subseteq$.}
As $Q$ is monotone, $Q(D') \subseteq Q(D)$ holds for any sub-database $D' \subseteq D$. 

%\underline{\em Direction $\supseteq$.} 
\vspace{-0.9\baselineskip}
\proofsubparagraph*{\boldmath Direction $\supseteq$.}
Consider an arbitrary query result $t \in Q(D)$. Let $D'_i(t)$ denote the group of tuples returned by $\ourprob(Q_i, \{R_j: R_j \in E_i\}, \pi_{\mathbf{A}_i} t)$. We note that $t \in \pi_{\mathbf{A}} \Join_{i \in [k]}D'_i(t)$, since 
\begin{itemize}
\item every tuple has the same value $\pi_A t$ over any output attribute $A$, if it contains attribute~$A$; 
\item there is no non-output attribute to join for tuples across groups;
\item tuples inside each group can be joined by non-output attribute; (implied by the correctness of  \ourprob\ for a single query result) 
\end{itemize}
Hence, $t \in Q(D')$. Together, $Q(D') \supseteq Q(D)$.

We next show that there exists no $D'' \subseteq D$ such that $Q(D'') = Q(D)$ and $|D''| < |D'|$. Suppose not, let $D''_i \subseteq D''$ denote the set of tuples from relations in $E_i$. As $|D''| < |D'|$, there must exist some $i \in [k]$ such that $|D''_i| < |D'_i|$, i.e., $D'_i - D''_i \neq \emptyset$. In Algorithm~\ref{alg:easy-SWP}, we can rewrite $D'_i$ as follows: 
\begin{align*}
D'_i 
& = \biguplus_{t' \in \pi_{\mathbf{A}_i} Q(D)} \ourprob(Q_i, \{R_i: R_i \in E_i\}, t'),
\end{align*}
where $\biguplus$ denotes the disjoint union. As $D'_i - D''_i \neq \emptyset$, there must exist some $t' \in \pi_{\mathbf{A}_i} Q(D)$ such that $t' \notin Q_i(D''_i)$, i.e., $t'$ cannot be witnessed by $D''_i$. Correspondingly, there must exist some query result $t$ with $\pi_{\mathbf{A}_i} t = t'$ such that $t \notin Q(D'')$, i.e. $t$ cannot be witnessed by $D''$, contradicting the fact that $Q(D'') = Q(D)$. Hence, no such $D''$ exists.
\end{proof}

%\subparagraph{Remark 1.}
\begin{remarknew}
\ourprob\ is poly-time solvable for any full CQ, since $\attr(R_i)  \cap \attr(R_j) \subseteq \head(Q)$ holds for every pair of relations $R_i,R_j \in \rel(Q)$. Hence, the hardness of \ourprob\ comes from projection. On the other hand, \ourprob\ is also poly-time solvable for some non-full CQs, say $Q(A_1, A_2, A_3) :- R_1(A_1,A_2), R_2(A_2, A_3),R_3(A_1,A_3), R_4(A_1, B_1)$.
\end{remarknew}

%\subparagraph{Remark 2.}
\begin{remarknew}
It is not always necessary to compute $Q(D)$ as Algorithm~\ref{alg:easy-SWP} does. We actually have much faster algorithms for some classes of CQs. If CQ $Q$ is full, \ourprob$(Q,D)$ is the set of {\em non-dangling} tuples in $D$, i.e., those participate in at least one query result of $Q(D)$. Furthermore, if $Q$ is an acyclic full CQ, non-dangling tuples can be identified in $O(|D|)$ time~\cite{yannakakis1981algorithms}. It is more expensive to identify non-dangling tuples for cyclic full CQs, for example, the PANDA algorithm~\cite{abo2016faq} can identify non-dangling tuples for any full CQ in $O(N^w)$ time, where $w \ge 1$ is the sub-modular width of input query~\cite{abo2016faq}. It is left as an interesting open question to compute \ourprob\ for CQs with head-cluster property more efficiently. 
\end{remarknew}

\subsection{Hardness}
\label{sec:exact-hardness}
We next prove the second part of Theorem~\ref{the:dichotomy-exact} by showing the hardness for CQs without head-cluster property. \rev{Our hardness result is built on the NP-hardness of set cover~\cite{bernhard2008combinatorial}: 
Given a universe $\mathcal{U}$ of $n$ elements and a family $\mathcal{S}$ of subsets of $\mathcal{U}$, it asks to find a subfamily ${\displaystyle {\mathcal {C}} \subseteq {\mathcal {S}}}$ of sets whose union is ${\displaystyle {\mathcal {U}}}$ (called ``cover''), while using the fewest sets.} We start with $\cover(A):-R_1(A, B), R_2(B)$ and then extend to any CQ without head-cluster property. 

\begin{lemma}
\label{lem:cover-hard}
{\ourprob\ is not poly-time solvable for  $\cover(A):-R_1(A, B), R_2(B)$, unless $\texttt{P} = \texttt{NP}$.}
\end{lemma}

\begin{proof}
We show a reduction from set cover to \ourprob\ for $\cover$. Given an arbitrary instance $(\mathcal{U}, \mathcal{S})$ of set cover, we construct a database $D$ for $\cover$ as follows. For each element $u \in \mathcal{U}$, we add a value $a_u$ to $\dom(A)$; for each subset $S \in \mathcal{S}$, we add a value $b_S$ to $\dom(B)$, and a tuple $(b_S)$ to $R_2$. Moreover, for each pair $(u, S) \in  \mathcal{U} \times \mathcal{S}$ with $u \in S$, we add a tuple $(a_u, b_S)$ to $R_1$. Note that $\cover(D)= \mathcal{U}$. 
It is now to show that $(\mathcal{U}, \mathcal{S})$ has a cover of size $\le k$ if and only if $\ourprob(\cover, D)$ has a solution $D'$ of size $\le |\mathcal{U}| + k$. Then, if \ourprob\ is poly-time solvable for $\cover$, set cover is also poly-time solvable, which is impossible unless $\texttt{P} = \texttt{NP}$.
\end{proof}

\begin{lemma}\label{lem:hardness}
For a CQ $Q$ without head-cluster property, \ourprob\ is not poly-time solvable for $Q$, unless $\texttt{P} = \texttt{NP}$.
\end{lemma}

\begin{proof}
Consider such a CQ $Q$ with a desired pair of relations $R_i, R_j \in \rel(Q)$. We next show a reduction from $\cover$ to $Q$. Given an arbitrary database $D_{\texttt{cover}}$ over $\cover$, we construct a database $D$ over $Q$ as follows. First, it is always feasible to identify attribute $A' \in \head(R_i) - \attr(R_j)$ and attribute $B' \in \attr(R_i) \cap \attr(R_j) - \head(Q)$. We set $\dom(A') = \dom(A)$, $\dom(B') = \dom(B)$, and remaining attributes with a dummy value $\{*\}$. 
Each relation in $Q$ degenerates to $R_1(A,B)$, or $R_2(B)$, or a dummy tuple $\{*,*,\cdots, *\}$. It can be easily checked that there is a one-to-one correspondence between solutions to $\ourprob(Q,D)$ and solutions to $\ourprob(\cover,D_{\texttt{cover}})$. Thus, if $\ourprob$ is poly-time solvable for $Q$, then $\ourprob$ is also poly-time solvable for $\cover$, coming to a contradiction of Lemma~\ref{lem:cover-hard}.
\end{proof}

\section{Dichotomy of Approximated SWP}
\label{sec:approximate}
As it is inherently difficult to compute \ourprob\ exactly for general CQs, the next interesting question is to explore approximated solutions for \ourprob. In this section, we establish the following dichotomy for approximating \ourprob. All missing proofs are given in Appendix~\ref{appendix:approximate}.

\begin{theorem}\label{the:dichotomy-approx}
If a CQ $Q$ has head-domination property, $\ourprob$ is $O(1)$-approximable; otherwise, $\ourprob$ of input size $N$ is not $(1 -o(1)) \cdot \log N$-approximable, unless $\texttt{P} = \texttt{NP}$.
\end{theorem}

\subsection{A \texorpdfstring{\boldmath $O(1)$}{O(1)}-Approximation Algorithm}
%\subsection{A $O(1)$-Approximation Algorithm}
\label{sec:approximate-easy}

Let's start by revisiting $\cover(A):-R_1(A,B), R_2(B)$.
Although \ourprob\ is hard to compute exactly for $\cover$, it is easy to approximate \ourprob$(\cover,D)$ for arbitrary database $D$ within a factor of $2$. Let $D^*$ be \rev{a} solution to \ourprob$(\cover, D)$. We can simply construct an approximated solution $D'$ by picking a pair of tuples $(a,b) \in R_1, (b) \in R_2$ for every $a \in \cover(D)$, and show that $|D'| \le 2 \cdot |\cover(D)| \le 2 \cdot |D^*|$. This is actually not a violation to the inapproximability of set cover problem. If revisiting the proof of Lemma~\ref{lem:cover-hard}, $(\mathcal{U}, \mathcal{S})$ has a cover of size $\le k$ if and only if $\ourprob(\cover, D)$ has a solution of size $\le |\mathcal{U}| + k$. Due to the fact that $k \le |\mathcal{U}|$, the inapproximability of set cover does not carry over to $\ourprob$ for $Q$. This observation can be generalized to all CQs with head-domination property. 

\enlargethispage{\baselineskip}
As described in Algorithm~\ref{alg:easy-SWP}, an approximated solution to \ourprob$(Q,D)$ for a CQ $Q$ with head-domination property consists of two parts. For every relation that only contains output attributes, Algorithm~\ref{alg:easy-SWP} includes all tuples that participate in at least one query result (line 4--5), which must be included by any witness for $Q(D)$. For the remaining relations, Algorithm~\ref{alg:easy-SWP} partitions them into groups based on the existential connectivity. Intuitively, every pair of relations across groups can only join via output attributes in their dominants. Recall that ${\bf A}_i$ denotes the set of output attributes appearing in relations from $E_i$. Then, for each group $E_i$, we consider each tuple $t' \in \pi_{{\bf A}_i}Q(D)$ and find the smallest witness for $t'$ in $Q_i$ defined by relations in $E_i$. The union of witnesses returned for all groups forms the final answer. As shown before, Algorithm~\ref{alg:easy-SWP} runs in polynomial time.

\rev{\begin{lemma}\label{lem:head-domination}
For a CQ $Q$ with head-domination property, Algorithm~\ref{alg:easy-SWP} finds a $O(1)$-approximated solution to \ourprob$(Q,D)$ for any database $D$ in polynomial time.
\end{lemma}}

\begin{proof}
\!Consider the connected components $E_1, E_2, \cdots, E_k$ of $G^\exists_Q$ with dominants $\dot{R}_1, \dot{R}_2, \cdots,$  $\dot{R}_k$. Let $Q_i$ be the subquery defined over relations in $E_i$, with output attributes $\head(Q_i) = \attr(\dot{R}_i)$, and $D_i = \{R_j : R_j \in E_i\}$ be the corresponding database for $Q_i$. Let $D^*$ be the solution to $\ourprob(Q,D)$. We point out some observations on $D^*$: 
\begin{itemize}
\item For each $R_j$ with $\attr(R_j) \subseteq \head(Q)$, $D^*$ must include all tuples in $\pi_{\attr(R_j)} Q(D)$. 
\item For every dominant $\dot{R}_i$, $D^*$ must include at least $|\pi_{\head(\dot{R}_i)} Q(D)|$ tuples from $\dot{R}_i$. 
\end{itemize} 

\begin{figure*}
\begin{minipage}{0.58\linewidth}
\centering\includegraphics[scale=1.1]{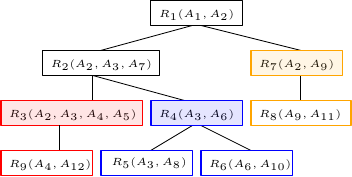}
\caption{A free-connex CQ $Q$ with $\head(Q)=\{A_1,A_2, A_3, A_7\}$. A partition of  relations containing non-output attributes is $\left\{\{R_3,R_9\}, \{R_4, R_5, R_6\}, \{R_7, R_8\}\right\}$, with dominant relations $R_3, R_4, R_7$ respectively.} 
\label{fig:free-connex}
\end{minipage}
\ \ 
\begin{minipage}{0.38\linewidth}
\centering
\includegraphics[scale=1.0]{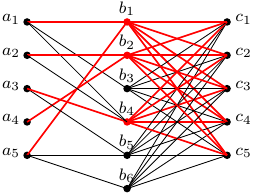}
\caption{A database $D$ for $\matrix$ with an integral sub-database in red. Each vertex is a value in the attribute, and each edge is a tuple in $D$.}
\label{fig:integral}   
\end{minipage}
\end{figure*}

On the other hand, Algorithm~\ref{alg:easy-SWP} includes $|\rel(Q_i)|$ tuples for each primitive at line 11, and invokes this primitive for each tuple in $\pi_{\head(\dot{R}_i)} Q(D)$. Together, we come to:
\begin{align*}
|D'| &= \sum_{R_j :\attr(R_j) \subseteq \head(Q)} \left|\pi_{\attr(R_j)} Q(D)\right| + \sum_{\dot{R}_i: i \in [k]} \left|\pi_{\head(\dot{R}_i)} Q(D)\right| \cdot |\rel(Q_i)|\\
&\le |D^*| + |D^*| \cdot |\rel(Q_i)| \le  2 \cdot |D^*| \cdot |\rel(Q)|
\end{align*}
It can also be easily checked that $Q(D) = Q(D')$.
Hence, Algorithm~\ref{alg:easy-SWP} always returns a $O(1)$-approximation solution for $\ourprob(Q,D)$.
Moreover, the query results $Q(D)$ can be computed in poly-time in the input size of $D$. Each primitive of finding the smallest witness for one query result takes $O(1)$ time. Hence, Algorithm~\ref{alg:easy-SWP} runs in polynomial time. 
\end{proof}
\subparagraph{Connection with Free-connex CQs.} We point out that every free-connex CQ has head-domination property, which is built on an important property as stated in Lemma~\ref{lem:free-connex-tree}.  

\begin{lemma}[\cite{bagan2007acyclic}]
\label{lem:free-connex-tree}
A free-connex CQ has a tree structure $\mathbb{T}$ such that (1) each node of $\mathbb{T}$ corresponds to $\attr(R_i)$ or $\head(R_i)$ for some relation $R_i \in \rel(Q)$; and for each relation $R_i \in \rel(Q)$, there exists a node of $\mathbb{T}$ corresponding to $\attr(R_i)$; (2) for every attribute $A \in \attr(Q)$, the set of nodes containing $A$ form a connected subtree of $\mathbb{T}$; (3) there is a connected subtree $\mathbb{T}_\textsf{con}$ of $\mathbb{T}$ including the root of $\mathbb{T}$, such that the set of attributes appearing in $\mathbb{T}_\textsf{con}$ is exactly $\head(Q)$.
\end{lemma} 

\begin{lemma}
\label{lem:free-connex-head-domination}
Every free-connex CQ has head-domination property.
\end{lemma}

\begin{proof}
Recall that $Q$ has head-domination property if for any connected component $E_j$ of $G^\exists_Q$, there exists a dominant relation from $\rel(Q)$ for $E_j$. Suppose we are given a tree structure $\mathbb{T}$ for a free-connex CQ $Q$ with $\mathbb{T}_\textsf{con}$ as described by Lemma~\ref{lem:free-connex-tree}. Consider an arbitrary relation $R_i$ with $\attr(R_i) - \head(R_i) \neq \emptyset$, such that all its ancestors in $\mathbb{T}$ only contain output attributes. Let $\mathbb{T}_i$ be the subtree rooted at $R_i$. We note that any relation in $\mathbb{T}_i$ cannot fall into the same connected component with a relation in $\mathbb{T} \setminus \mathbb{T}_i$, since they do not share any common non-output attribute. Consider any relation $R_j \in \mathbb{T}_i$. Implied by the fact that $\mathbb{T}_\textsf{con}$ is a connected subtree and $R_i \notin \mathbb{T}_\textsf{con}$, we have $R_j \notin \mathbb{T}_\textsf{con}$. Then, $\head(R_j) \subseteq \head(R_i)$; otherwise, any output attribute in $\head(R_j) - \head(R_i)$ does not appear in $\mathbb{T}_\textsf{con}$. Hence, for any connected component $E_j$ formed by relations from $\mathbb{T}_i$, $R_i$ is a dominant relation for $E_j$. This argument applies for every connected component of $G^\exists_Q$.
\end{proof}

\subsection{Logarithmic Inapproximability}
\label{sec:approximate-hard}

Now, we turn to the class of CQs without head-domination property, and show their hardness by resorting to inapproximability of set cover~\cite{feige1998threshold, dinur2014analytical}: there is no poly-time algorithm for approximating set cover of input size $n$ within factor $(1 -o(1)) \cdot \log n$, unless \texttt{P} ${\displaystyle =}$ \texttt{NP}. We identify two hardcore structures (Definition~\ref{def:free-sequence} and  Definition~\ref{def:nested-clique}), and prove that no poly-time algorithm can approximate \ourprob\ for any CQ containing a hardcore within a logarithmic factor, unless $\texttt{P} = \texttt{NP}$. Lastly, we complete the proof of Theorem~\ref{the:dichotomy-approx} by establishing the connection between the non-existence of a hardcore and head-domination property for CQs.

\subsubsection{Free sequence}
Let's start with the simplest acyclic but non-free-connex CQ  {$\matrix(A,C) :\!- R_1(A,B), R_2(B,\\C)$}. We show a reduction from set cover to $\ourprob$ for $\matrix$ while preserving its logarithmic inapproximability. The essence of is the notion of {\em integral} witness, such that for any database $D$ where $R_2$ is a Cartesian product, $\ourprob(\matrix, D)$ always admits an integral witness. 

\begin{definition}[Integral Database]
For any database $D$ over $\matrix$ with $R_2 = (\pi_B R_2) \times (\pi_C R_2)$, a sub-database $(R'_1,R'_2) \subseteq D$ is integral if $R'_2 = (\pi_{B} R'_2) \times (\pi_C R_2)$. 
\end{definition}

\begin{lemma}
\label{lem:matrix-integral}
For $\matrix$ and any database $D$ {where $R_2 = (\pi_B R_2) \times (\pi_C R_2)$}, there is an integral solution to $\ourprob(\matrix, D)$, i.e., a smallest witness to $Q(D)$ that is also integral. 
\end{lemma}
\begin{figure}[t]
\end{figure}
{\begin{proof}
Consider a database $D$ where $R_2 = (\pi_B R_2) \times (\pi_C R_2)$. We assume that there exists no dangling tuples in $R_1, R_2$, i.e., every tuple can join with some tuple from the other relation; otherwise, we simply remove these dangling tuples. With a slight abuse of notation, we denote $A = \pi_A R_1$, $B = \pi_B R_1$($= \pi_B R_2$) and $C = \pi_C R_2$. We consider the optimization problem $O1$ in Figure~\ref{fig:optimization}.  
Intuitively, it asks to assign a subset of elements $C_b \subseteq C$ to each value $b \in B$, such that each $a$ is ``connected'' to all values in $C$ via tuples in $R_1, R_2$, while the total size of the assignment defined as $\sum_{b \in B} |C_b|$ is minimized. See Figure~\ref{fig:integral}.

We rewrite the objective function as:
$\sum_{b \in B} |C_b| = \sum_{c \in C} \left|\{b \in B: \\c \in C_b\}\right|$, where $\{b \in B: c \in C_b\}$ indicates the subset of values from $B$ to which $c$ is assigned. Together with the constraint that every $a \in A$ must be connected to $c$, we note that minimizing $|\{b \in B: c \in C_b\}|$ is equivalent to solving the optimization problem $(O2)$ in Figure~\ref{fig:optimization}.  
Let $x^*$ be the optimal solution of the program above, which only depends on the input relation $R_1$, and completely independent of the specific value $c$. Hence, we conclude that 
$\sum_{b \in B} |C_b|=|C| \cdot \sum_{b \in B} x^*_b$.
\begin{figure}
\centering
\minipage{0.25\textwidth}
\begin{align*}
(O1) \ \ \ \ \ \min & \sum_{b \in B} |C_b| \ \ \ \ \ \ \\
\textrm{s.t.} \ & \bigcup_{b: (a,b) \in R_1} C_b = C, \forall a \in A \\ 
& C_b \subseteq C, \forall b \in B
\end{align*}
\endminipage
\ \ \ \ \ \ \ \ \ 
\minipage{0.25\textwidth}
\begin{align*}
(O2) \ \ \ \ \ \min & \  \sum_{b \in B} x_b \ \ \ \ \ \\
\textrm{s.t.} \ & \sum_{b: (a, b) \in R_1} x_b \ge 1, \forall a \in A \\
& x_b \in \{0,1\}, \forall b \in B
\end{align*}
\endminipage
\vspace{-1em}
\caption{Optimization problems in the proof of Lemma~\ref{lem:matrix-integral}.}
\label{fig:optimization}
\end{figure}

We next construct an integral sub-database $D'$ as follows. For each $b \in B$ with $x^*_b =1$, we add tuples in $\{(b,c) \in R_2: \forall c \in C\}$ to $D'$. For each $a \in A$, we pick an arbitrary $b \in B$ with $(a,b) \in R_1$ and $x^*_b =1$, and add $(a,b)$ to $D'$. Any solution to $\ourprob(\matrix, D)$ must contain at least $|A|$ tuples from $R_1$ and at least $|C| \cdot \sum_{b \in B} x^*_b$ tuples from $R_2$. Hence, $D'$ is an integral witness to $\ourprob(\matrix, D)$. 
\end{proof}}

\begin{lemma}
\label{lem:matrix}
There is no poly-time algorithm to approximate \ourprob\ for $\matrix$ within a factor of $(1-o(1)) \cdot \log N$, unless \texttt{P} ${\displaystyle =}$ \texttt{NP}.
\end{lemma}

% custom setting footnote margin
\addtolength{\skip\footins}{-10pt}
\begin{proof}
\enlargethispage{1.6\baselineskip}
Consider an arbitrary instance of set cover $(\mathcal{U},\mathcal{S})$, where $|\mathcal{U}| = n$ and $|\mathcal{S}| = n^c$ for some constant $c \ge 1$.\footnote{The inapproximability of set cover holds even when the size of the family of subsets is only polynomially large with respect to the size of the universe of elements~\cite{moshkovitz2012projection}.} 
We construct a database $D$ for $\matrix$ as follows. For each element $u \in \mathcal{U}$, we add a value $a_u$ to $\dom(A)$ and $c_u$ to $\dom(C)$; for each subset $S \in \mathcal{S}$, we add a value $b_S$ to $\dom(B)$. Moreover, for each pair $(u, S) \in  \mathcal{U} \times \mathcal{S}$ with $u \in S$, we add tuple $(a_u, b_S)$ to $R_1$.  \rev{$R_2$ is a Cartesian product between $\dom(B)$ and $\dom(C)$.} Every relation contains at most $n^{c+1}$ tuples. Hence $N \le 2 n^{c+1}$. Note that $\matrix(D)$ is the Cartesian product between $A$ and $C$.
Implied by Lemma~\ref{lem:matrix-integral}, it suffices to consider integral witness to $\ourprob(\matrix, D)$. Here, we show that $(\mathcal{U}, \mathcal{S})$ has a cover of size $\le k$ if and only if the $\ourprob(\matrix, D)$ has an integral solution of size $\le |\mathcal{U}| + k \cdot |V| = n(k+1)$.

%\underline{\em Direction only-if.}
\vspace{-0.9\baselineskip}
\proofsubparagraph*{Direction only-if.}
Suppose we are given a cover $\mathcal{S}'$ of size $k$ to $(\mathcal{U}, \mathcal{S})$. We construct an integral solution $D'$ to $\ourprob(\matrix, D)$ as follows. Let $B' \subseteq \dom(B)$ be the set of values that corresponding to $\mathcal{S}'$. For every $b_S \in B'$, i.e., $S \in \mathcal{S}'$, we add tuple $(b_S, c_u)$ to $D'$ for every $u \in \mathcal{U}$. For every $a_u \in \dom(A)$, we choose an arbitrary value $b_S \in B'$ such that $u \in S$, and add tuple $(a_u, b_S)$ to $D'$. This is always feasible since $\mathcal{S}'$ is a valid set cover. It can be easily checked that $n$ tuples from $R_1$ and $k \cdot n$ tuples from $R_2$ are added to $D'$.
%\underline{\em Direction if.}
\vspace{-0.9\baselineskip}
\proofsubparagraph*{Direction if.}
Suppose we are given a integral solution $D'$ of size $k'$ to $\ourprob(\matrix, D)$. Let $B'$ be the subset of values whose incident tuples in $R_2$ are included by $D'$. We argue that all subsets corresponding to $B'$ forms a valid cover of size $\frac{k'}{n-1}$. By definition of integral solution, $|B'| = \frac{k'}{n} -1$. Moreover, for every value $a \in \dom(A)$, at least one edge $(a,b)$ is included by $D'$ for some $b \in B'$, hence $B'$ must be a valid set cover. 

Hence, if $\ourprob$ is $(1-o(1)) \cdot \log N$-approximable for $\matrix$, there is a poly-time algorithm that approximates set cover of input size $n$ within a $(1-o(1)) \cdot \log N$ factor, which is impossible unless \texttt{P} ${\displaystyle =}$ \texttt{NP}.
\end{proof}

\begin{definition}[Free Sequence]
\label{def:free-sequence}
In a CQ $Q$, a free sequence is a sequence of attributes $P=\langle A_1, A_2, \cdots, A_k\rangle$ such that\footnote{Free sequence is a slight generalized notion of free path~\cite{bagan2007acyclic} studied in the literature, which further requires that for any relation $R_j \in \rel(Q)$, either $\attr(R_j) \cap P = \emptyset$, or $|\attr(R_j) \cap P |=1$, or $\attr(R_j) \cap P = \{A_i, A_{i+1}\}$ for some $i \in [k-1]$. }
\begin{itemize}
\item $A_1, A_k \in \head(Q)$ and $A_2, A_3, \cdots, A_{k-1} \in \attr(Q) - \head(Q)$;
\item for every $i \in [k-1]$, there exists a relation $R_j \in \rel(Q)$ such that $A_i, A_{i+1} \in \attr(R_j)$;
\item there exists no relation $R_j \in \rel(Q)$ such that $A_1, A_k \in R_j$.
\end{itemize}
\end{definition}

\begin{lemma}
\label{lem:free-sequence}
For a CQ $Q$ containing a free sequence, there is no poly-time algorithm that can approximate $\ourprob$ for $Q$ within a factor of $(1-o(1)) \cdot \log N$, unless \texttt{P} ${\displaystyle =}$ \texttt{NP}.
\end{lemma}

\enlargethispage{2\baselineskip}
\begin{proof}
Let $P = \langle A_1, A_2, \cdots, A_k\rangle$ be such a sequence. For simplicity, let $P' = \{A_2, A_3, \cdots,$ $A_{k-1}\}$. We next show a reduction from set cover to $\ourprob(Q,D)$. Consider an arbitrary instance of set cover with a universe $\mathcal{U}$ and a family $\mathcal{S}$ of subsets of $\mathcal{U}$ where $|\mathcal{U}| = n$ and $|\mathcal{S}| = n^c$ for some constant $c \ge 1$. We construct a database $D$ for $Q$ as follows. For each $u \in \mathcal{U}$, we add a value $a_u$ to $\dom(A_1)$ and $\dom(A_k)$. For each subset $S \in \mathcal{S}$, we add a value $b_S$ to $\dom(B)$ for every $B \in P'$. We set the domain of remaining attributes in $\attr(Q) - P$ as $\{*\}$. For each relation $R_j \in \rel(Q)$, we distinguish the following cases:
\begin{itemize}
\item If $A_1 \in \attr(R_j)$, we further distinguish two more cases: 
\begin{itemize}
\item if $\attr(R_j) \cap P' = \emptyset$, we add tuple $t$ with $\pi_{A_1} t = a_u$ for every $u \in \mathcal{U}$;
\item otherwise, we add tuple $t$ with $\pi_{A_1} t = a_u$ and $\pi_{A_i} t = b_S$ for $A_i \in \attr(R_j) \cap P'$, for every pair $(u,S) \in \mathcal{U} \times \mathcal{S}$ such that $u \in S$;
\end{itemize}
\item If $A_k \in \attr(R_j)$, we further distinguish two more cases: 
\begin{itemize}
\item if $\attr(R_j) \cap P' = \emptyset$, we add tuple $t$ with $\pi_{A_k} t = a_u$ for every $u \in \mathcal{U}$;
\item otherwise, we add tuple $t$ with $\pi_{A_k} t = a_u$ and $\pi_{A_i} t = b_S$ for $A_i \in \attr(R_j) \cap P'$, for every pair $(u,S) \in \mathcal{U} \times \mathcal{S}$;
\end{itemize}
\item If $P \cap \attr(R_j) \subseteq P - \{ A_1, A_k\}$, we add a tuple $t$ with $\pi_{A_i} t = b_S$ for $A_i \in \attr(R_j) \cap P$, for every $S \in \mathcal{S}$;
\item If $P \cap \attr(R_j) = \emptyset$, then we add a tuple $\{*\}$;
\end{itemize} 
It can be easily checked that every relation contains at most $n^{c+1}$ tuples,  hence $\log N= \Theta(\log n)$. The query result $Q(D)$ is exactly the Cartesian product of $\mathcal{U} \times \mathcal{U}$.

Consider a sub-database $D'$ of $D$ constructed above. Let $R'_j$ be the corresponding sub-relation of $R_j$ in $D'$. A solution $D'$ to \ourprob$(Q,D)$ is integral if $R'_j = \left(\pi_{\attr(R_j) \cap P'} R'_j\right) \times \left(\pi_{A_k} R_j\right)$ holds for every relation $R_j$ with $A_k \in \attr(R_j)$ and $\attr(R_j) \cap P' \neq \emptyset$. Applying a similar argument as Lemma~\ref{lem:matrix-integral}, we can show that there always exists an integral solution to \ourprob$(Q,D)$. Below, it suffices to focus on integral solutions. 
It can be easily proved that $(\mathcal{U},\mathcal{S})$ has a cover of size $\le k$ if and only if \ourprob$(Q,D)$ has an integral solution of size $\le
n q_1+ kn q_2 + q_3$ where $q_1,q_2,q_3 \le |\rel(Q)|$ are query-dependent parameters.  If $\ourprob$ is $(1-o(1)) \cdot \log N$-approximatable for $Q$, there is a poly-time algorithm that can approximate set cover instances of input size $n$ within a $\log n$-factor, which is impossible unless $\texttt{P} = \texttt{NP}$. 
\end{proof}

\subparagraph{Connection with Non-Free-connex CQs.} We point out that every acyclic but non-free-connex CQ has a free sequence~\cite{bagan2007acyclic}, hence does not have head-domination property. Together with Lemma~\ref{lem:free-connex-head-domination}, our characterization of \ourprob\ for acyclic CQs coincides with the separation between free-connex and non-free-connex CQs. In short, \ourprob\ is poly-time solvable or $O(1)$-approximable for free-connex CQs, while no poly-time algorithm can approximate \ourprob\ for any acyclic but non-free-connex CQs within a factor of $(1-o(1)) \cdot \log N$, unless \texttt{P} ${\displaystyle =}$ \texttt{NP}.

\subsubsection{Nested Clique}
Although free sequence suffices to capture the hardness of approximating \ourprob\ for acyclic CQs, it is not enough for cyclic CQs. Let's start with the simplest cyclic CQ $\pyramid(A,B,C):-R_1(A,B), R_2(A,C), R_3(B,C), R_4(A,F), R_5(B,F), R_6(C,F)$ that does not contain a free sequence, but $\ourprob$ is still difficult to approximate.

\begin{lemma}
\label{lem:pyramid}
There is no poly-time algorithm to approximate $\ourprob$ for $\pyramid$ within a factor of $(1-o(1)) \cdot \log N$, unless \texttt{P} ${\displaystyle =}$ \texttt{NP}.
\end{lemma}
%\enlargethispage{\baselineskip}
\begin{proof}
Consider an instance $(\mathcal{U},\mathcal{S})$ of set cover, with $\mathcal{U} = \{u_1,u_2, \cdots, u_n\}$ and $\mathcal{S} = \{S_1,S_2, \cdots, S_m\}$, where $m = n^c$ for some constant $c >1$. We construct a database $D$ for $Q$ as follows. Let $\dom(A) = \{a_1,a_2, \cdots, a_n\}$, $ \dom(B) = \{b_1, b_2, \cdots, b_n\}$, $\dom(C) = \{c_1, c_2, \cdots, c_n\}$ and $\dom(F) = \dom(F^-) \times \dom(F^+)$, where $\dom(F^-)=\{f^-_1, f^-_2, \cdots, f^-_m\}$ and $\dom(F^+)=\{f^+_1,f^+_2, 
\cdots, f^+_n\}$. Relations $R_1$, $R_2$, $R_3$ and $R_6$ are Cartesian products of their corresponding attributes. For each pair $(u_\ell, S_j) \in \mathcal{U} \times \mathcal{S}$ with $u_\ell \in S_j$, we add tuples of $\{(a_\ell, f^-_j)\} \times \dom(F^+)$ to $R_4$. For each $i \in [n]$, we add tuples $\{b_i\} \times \dom(F^-) \times \{f^+_i\}$ to $R_5$. It can be easily checked that the input size of $D$ is $O(n^{2c})$, hence $\log N = \Theta(\log n)$. $Q(D)$ is the Cartesian product between $A$,$B$ and $C$. Hence, every solution to $\ourprob(Q,D)$ includes all tuples in $R_1, R_2, R_3$. Below, we focus on $R_4, R_5, R_6$.

We observe that $D$ enjoys highly symmetric structure over $\dom(B)$. More specifically, each value $b_i \in \dom(B)$ induces a subquery 
$Q_i(A,C) = R_4(A,F_i) \Join R_5(F_i) \Join R_6(C,F_i)$,
where $\dom(F_i) = \dom(F^-) \times \{f^+_i\}$, and a sub-database $D^i = \left\{R^i_4, R^i_5, \\ R^i_6\right\}$, where $R^i_4 = \{(a_\ell, f^-_j, f^+_i): \forall \ell \in [n], j \in [m], u_\ell \in S_j\}$, $R^i_5 = \dom(F_i)$ and $R^i_6 = \dom(C) \times \dom(F_i)$. It can be easily checked that 
$\ourprob(Q,D) = R_1 \uplus R_2 \uplus R_3 \uplus \left(\uplus_{i \in [n]}\ourprob(Q_i,D_i)\right)$.
For any $i \in [n]$, computing $\ourprob(Q_i,D_i)$ is almost the same as $\matrix$. Moreover, the solution to each $\ourprob(Q_i,D_i)$ shares the same structure, which is independent of the specific value $b_i \in \dom(B)$. In a sub-database $D' \subseteq D$, let $R'_i$ be the corresponding sub-relation of $R_i$. A solution $D'$ to $\ourprob(Q,D)$ is {\em integral} if $R'_4 = (\pi_{A, F^-} R'_4) \times \dom(F^+)$, $R'_5 = \dom(B) \times \left(\pi_{F^-} R'_5\right) \times \dom(F^+)$, and $R'_6 = \dom(C) \times \left(\pi_{F^-} R'_6\right) \times \dom(F^+)$. Implied by Lemma~\ref{lem:matrix-integral} and analysis above, there always exists an integral solution to $\ourprob(Q, D)$.

Moreover, $(\mathcal{U}, \mathcal{S})$ has a cover of size $\le k$ if and only if $\ourprob(Q, D)$ has an integral witness of size $\le 3n^2 + n(n+k+kn) = (k+4)n^2 + kn$. If $\ourprob$ is $(1-o(1)) \cdot \log N$-approximable for $Q$, then there is a poly-time algorithm that can approximate set cover instances of input size $n$ within a factor of $(1-o(1)) \cdot \log n$, which is impossible unless \texttt{P} ${\displaystyle =}$ \texttt{NP}.
\end{proof}

Now, we are ready to introduce the structure of {\em nested clique} and the {\em rename} procedure for capturing the hardness of cyclic CQs:

\begin{definition}[Nested Clique]
\label{def:nested-clique}
In a CQ $Q$, a nested clique is a subset of attributes $P \subseteq \attr(Q)$ such that 
\begin{itemize} 
\item for any pair of attributes $A,B \in P$, there is some $R_j \in \rel(Q)$ with $A,B \in \attr(R_j)$; 
\item $P \cap \head(Q) \neq \emptyset$ and $P - \head(Q) \neq \emptyset$;
\item there is no relation $R_j \in \rel(Q)$ with $P \cap \head(Q) \subseteq \head(R_j)$.
\end{itemize} 
\end{definition}

\begin{definition}[Rename]
Given the nonout-connectivity graph $H_Q$ of a CQ $Q$ with connected components $H_1,H_2,\cdots, H_k$, the rename procedure assigns one distinct attribute to all attributes in the same component. The resulted CQ $Q'$ contains the same output attributes as $Q$, and each $R_i \in \rel(Q)$ defines a new relation $R'_i \in \rel(Q')$ with $\attr(R'_i) = \head(R_i) \cup \{F_j:\forall j \in [k], H_j \cap \attr(R_i) \neq \emptyset\}$. 
\end{definition}

In Figure~\ref{fig:connectivity}, $Q_1$ is renamed as $Q'_1(A_1,A_2,A_3):-R_1(A_1,F_1), R_2(F_1), R_3(A_2,F_1)$, \\ $R_4(A_2,A_3,F_2),$ $R_5(A_1,A_2)$.

\begin{theorem}
\label{the:nested-clique}
For a CQ $Q$, if its renamed query $Q'$ contains a nested clique, there is no poly-time algorithm that can approximate $\ourprob$ for $Q$ within a factor of $(1-o(1)) \cdot \log N$, unless \texttt{P} ${\displaystyle =}$ \texttt{NP}.
\end{theorem}

\begin{proof}
Let $P \subseteq \attr(Q)$ be the subset of attributes corresponding to the clique in the renamed query of $Q$.
We identify the relation that contains the most number of output attributes of $P$, say $R_2 = \arg \max_{R_i \in \rel(Q)} |\attr(R_i) \cap P \cap \head(Q)|$. As there exists no relation $R_k \in \rel(Q)$ with $\head(Q) \cap P \subseteq \attr(R_k)$, it is always feasible to identify an attribute $A \in \head(Q) \cap P - \attr(R_2)$. For simplicity, we denote $P' = \head(Q) \cap P - \attr(R_2) - \{A\} = \{A_1, A_2, \cdots, A_\ell\}$, for some integer $\ell$.
All non-output attributes in $P - \head(Q)$ collapse to a single attribute $F$. All other attributes in $\attr(Q) -P$ contain a dummy value $\{*\}$.

Consider an arbitrary instance of set cover $(\mathcal{U},\mathcal{S})$ with $\mathcal{U} = \{u_1,u_2, \cdots, u_n\}$ and $\mathcal{S} = \{S_1,S_2, \cdots, S_m\}$, where $m = n^c$ for some constant $c >1$. We construct a database $D$ for $Q$. Let $\dom(A) = \{a_1,a_2, \cdots, a_n\}$, $ \dom(B) = \{b_1, b_2, \cdots, b_n\}$, $\dom(C) = \{c_1, c_2, \cdots, c_n\}$ for every $C \in P'$, and $\dom(F) = \dom(F^-) \times \dom(F^+)$ where $\dom(F^-) = \{f^-_1,f^-_2, \cdots, f^-_m\}$ and $\dom(F^+) = \{f^+_{h_1,h_2,\cdots, h_\ell}: \forall h_1, h_2,\cdots, h_\ell \in [n]\}$. We distinguish the following cases:
\begin{itemize}
\item If $\attr(R_i) \cap P \subseteq \head(Q)$, $R_i$ is a Cartesian product over all attributes in $\head(R_i) \cap P$;
\item Otherwise, we further distinguish the following three cases:
\begin{itemize}
\item $R_2$ is a Cartesian product over all attributes in $\attr(R_2) \cap P$;
\item If $A,F \in \attr(R_i)$, we construct sub-relation $\left(\pi_{A,F} R_i\right)$ such that for each pair $(u_\ell, S_j) \in \mathcal{U} \times \mathcal{S}$ with $u_\ell \in S_j$, we add tuples $\left\{a_\ell, f^-_j\right\} \times \dom(F^+)$ to $\left(\pi_{A,F} R_i\right)$. If $\attr(R_i) \cap P' \neq \emptyset$, for each tuple $\left(a_\ell, f^+_{h_1,h_2,\cdots,h_\ell}, f^-_j\right) \in \left(\pi_{A,F} R_i\right)$, we extend it by attaching value $c_{h_j}$ for attribute $A_j \in \attr(R_i) \cap P'$. This way, we already obtain $\left(\pi_{A,F, \attr(R_i) \cap P'} R_i\right)$. At last, we construct $R_i$ as the Cartesian product of remaining attributes in $\head(R_i) \cap \head(R_2) \cap P$ and $\left(\pi_{A,F, \attr(R_i) \cap P'} R_i\right)$.
\item Otherwise, $F \in \attr(R_i)$ but $A \notin \attr(R_i)$. We construct sub-relation $\left(\pi_{F} R_i\right) = \dom(F)$. If $\attr(R_i) \cap P' \neq \emptyset$, for each tuple $\left(f^+_{h_1,h_2,\cdots,h_\ell}, f^-_j\right) \in \pi_{F} R_i$, we extend it by attaching value $c_{h_j}$ for attribute $A_j \in \attr(R_i) \cap P'$. This way, we already obtain $\left(\pi_{F, \attr(R_i) \cap P'} R_i\right)$. At last, we construct $R_i$ as the Cartesian product of remaining attributes in $\head(R_i) \cap \head(R_2) \cap P$ and $\left(\pi_{A,F, \attr(R_i) \cap P'} R_i\right)$.
\end{itemize}
\end{itemize}
It can be checked that every relation contains $O(n^{|\head(Q) \cap P|} \cdot m)$ tuples, hence $\log N = \Theta(\log n)$. Meanwhile, the query result $Q(D)$ is the Cartesian product over attributes in $\head(Q) \cap P$, so every solution to $\ourprob(Q,D)$ must contain all tuples in $R_i$ if $\attr(R_i) \cap P \subseteq \head(Q)$, and the dummy tuple $\{*\}$ in every relation $R_i$ if $\attr(R_i) \cap P = \emptyset$.

% set back footnote margin
\addtolength{\skip\footins}{10pt}
Here, $D$ also enjoys highly symmetric structure over every attribute $C \in P'$. More specifically, every tuple $t = (c_{i_1}, c_{i_2}, \cdots, c_{i_\ell}) \in \times_{C \in P'} \dom(C)$ induces a subquery by removing all attributes in $P'$, and restricting $\dom(F)$ as $\dom(F_t) = \dom(F^-) \times \left\{f^+_{i_1},f^+_{i_2}, \cdots, f^+_{i_\ell}\right\}$. In a sub-database $D' \subseteq D$, let $R'_i$ be the corresponding sub-relation $R_i$. A solution $D'$ to $\ourprob(Q,D)$ is {\em integral} if:
\begin{itemize}
\item for every relation $R_i \in \rel(Q)$ with $A,F \in \attr(R_i)$, 
\begin{align*}
&\pi_{A,F, \head(R_i) \cap \head(R_2) \cap P} R'_i =\! \left(\pi_{A,F^-} R'_i\right) \times \dom(F^+) \times\! \left(\times_{C \in \head(R_i) \cap \head(R_2) \cap P} \dom(C)\right)
\end{align*}
\item for every relation $R_i \in \rel(Q)$ with $A \notin \attr(R_i)$ and $F \in \attr(R_i)$, 
\begin{align*}
& \pi_{F,\head(R_i) \cap \head(R_2) \cap P} R'_i = \left(\pi_{F^-} R'_i\right) \times \dom(F^+) \times \left(\times_{C \in \head(R_i) \cap \head(R_2) \cap P} \dom(C)\right)
\end{align*}
\end{itemize} 
It can be easily shown that there always exists an integral solution to $\ourprob(Q,D)$. Moreover, $(\mathcal{U}, \mathcal{S})$ has a cover of size $\le k$ if and only if $\ourprob(Q,D)$ has an integral solution of size $(q_1 k + q_2) \cdot n^{|\head(Q)\cap P|-1}$, where $q_1, q_2 \le |\rel(Q)|$ are some query-dependent parameters. Applying a similar argument as Lemma~\ref{lem:matrix-integral}, 
we can show that if $\ourprob$ is $(1-o(1)) \cdot \log N$-approximable for $Q$, there is a poly-time algorithm that can approximate set cover instances of input size $n$ within a $\log n$-factor, which is impossible unless $\texttt{P} = \texttt{NP}$. 
\end{proof}

\subsection{Completeness}
\label{sec:completeness}

At last, we complete the proof of Theorem~\ref{the:dichotomy-approx} by establishing the connection between the non-existence of hardcore structures and head-domination property:

\begin{lemma}
\label{lem:connection}
In a CQ $Q$, if there is neither a free sequence nor a nested clique in its renamed query, $Q$ has head-domination property.
\end{lemma}

We have proved Lemma~\ref{lem:connection} for acyclic CQs, such that if $Q$ does not contain a free sequence, $Q$ has head-cluster property. It suffices to focus on cyclic CQs. We note that any cyclic CQ contains a {\em cycle} or {\em non-conformal clique}~\cite{brault2016hypergraph}. Our proof is based on a technical lemma that if every cycle or non-conformal clique contains only output attributes or only non-output attributes, $Q$ has head-domination property. We complete it by showing that if $Q$ does not contain a free sequence or nested clique in its renamed query, no cycle or non-conformal clique contains both output and non-output attributes. 

\section{Approximation Algorithms}
\label{sec:approximation-algorithm}

We next explore possible approximation algorithms for CQs without head-domination property. All missing proofs are in Appendix~\ref{appendix:approximation-algorithm}. 

\subsection{Baseline}
\label{sec:baseline}
A baseline for general CQs returns the union of smallest witness for every query result, which includes at most $\min\{N, |\rel(Q)| \cdot |Q(D)|\}$ tuples. Meanwhile, AGM bound~\cite{AGM2008} implies that at least $O(|Q(D)|^{1/\rho^*})$ tuples are needed to reproduce $|Q(D)|$ results, where $\rho^*$ is the fractional edge covering number of $Q$. Together, we obtain: 

\begin{theorem}
\label{the:baseline}
\ourprob\ is $N^{1-1/\rho^*}$-approximable for any CQ $Q$, where $\rho^*$ is the fractional edge covering number of $Q$.
\end{theorem}

This upper bound is polynomially larger than the logarithmic lower bound proved in Section~\ref{sec:approximate}. We next explore better approximation algorithms for some commonly-used CQs. 

\subsection{Star \texorpdfstring{\rev{CQs}}{CQs}}
\label{sec:star}

We look into one commonly-used class of CQs noted as {\em star} CQs: \[Q_\textrm{star}(A_1, A_2, \cdots, A_m):- R_1(A_1,B), R_2(A_2, B), \cdots, R_m(A_m,B).\] Our approximation algorithm follows the greedy strategy developed for weighted set cover problem, where each element to be covered is a query result and each subset is a collection of tuples. Intuitively, the greedy strategy always picks a subset that minimizes the ``price'' for covering the remaining uncovered elements. The question boils down to specifying the universe $\mathcal{U}$ of elements, and the family $\mathcal{S}$ of subsets as well as their weights. However, naively taking every possible sub-collection of tuples from the input database as a subset, would generate an exponentially large $\mathcal{S}$, which leads to a greedy algorithm running in exponential time. Hence, it is critical to keep the size of  $\mathcal{S}$ small. Let's start with $\matrix$ ($m=2$). 

\subparagraph{Greedy Algorithm for \texorpdfstring{\boldmath $\matrix$}{\{Q\_matrix\}}.}
%\subparagraph{Greedy Algorithm for $\matrix$}  
Given $\matrix$, a database $D$, and a subset of query results $\mathcal{C} \subseteq \matrix(D)$, the price of a collections of tuples $(X,Y)$ for $X \subseteq R_1$ and $Y \subseteq R_2$ is defined as 
$f(\mathcal{C}, X, Y) = \frac{|X| + |Y|}{|\pi_{A,C}(X \Join Y) - \mathcal{C}|}$.
To shrink the space of candidate subsets, a critical observation is that the subset with minimum price chosen by the greedy algorithm cannot be {\em divided} further, i.e., all tuples should have the same join value as captured by Lemma~\ref{lem:decomposable}. 
\begin{lemma}
\label{lem:decomposable}
Given $\matrix$ and a database $D$,
for two distinct values $b,b' \in \dom(B)$, and two pairs of subsets of tuples $(X_1, Y_1) \in 2^{\sigma_{B=b}R_1} \times 2^{\sigma_{B=b}R_2}, (X_2, Y_2) \in 2^{\sigma_{B=b'} R_1} \times 2^{\sigma_{B=b'}R_2}$, for arbitrary $\mathcal{C} \subseteq \matrix(D)$, $\min\{ f(\mathcal{C}, X_1, Y_1), f(\mathcal{C}, X_2, Y_2)\} \le f(\mathcal{C}, X_1 \cup X_2, Y_1 \cup Y_2)$.
\end{lemma}

Now, we are ready to present a greedy algorithm that runs in polynomial time. As shown in Algorithm~\ref{alg:greedy-matrix}, we always maintain a pair $(X_i, Y_i)$ for every value $b_i \in \dom(B)$, such that $(X_i,Y_i)$ minimize the function $f(\mathcal{C}, X, Y)$ for $X \subseteq \sigma_{B = b_i} R_1$ and $Y \subseteq \sigma_{B=b_i} R_2$.
%Initially when $\mathcal{C} = \emptyset$, it can be easily proved that when $\mathcal{C} = \emptyset$, the pair $(X,Y)$ that leads to minimum price must be $X= \pi_A \sigma_{B= b_i}R_1$ and $Y = \pi_C \sigma_{B = b_i}R_2$. 
The greedy algorithm always chooses the one pair with minimum price, pick all related tuples in this pair, and update the coverage $\mathcal{C}$. After this step, we also need to update the candidate pair $(X_i, Y_i)$ for each value $b_i \in \dom(B)$ and enter into the next iteration. We will stop until all query results are covered.
\begin{algorithm}[t]
\caption{{\sc GreedySWP}$(\matrix,D)$.}
\label{alg:greedy-matrix}
$D' \gets \emptyset$, $\mathcal{C} \gets \emptyset$\;
\lForEach{$b_i \in \dom(B)$}{
$(X_i,Y_i) \gets (\sigma_{B=b_i} R_1, \sigma_{B=b_i} R_2)$}
\While{$\mathcal{C} \neq \matrix(D)$}{
$b_j \gets \arg \min_{b_i \in \dom(B)} f(\mathcal{C}, X_i, Y_i)$\;
$D' \gets D' \cup X_j \cup Y_j$, $\mathcal{C} \gets \mathcal{C} \cup \pi_{A,C}(X_j \Join Y_j)$\;
\ForEach{$b_i \in \dom(B)$}{
$\displaystyle{(X_i, Y_i) \gets \arg \min_{X \subseteq \sigma_{B=b_i} R_1, Y \subseteq \sigma_{B =b_i} R_2} f(\mathcal{C}, X, Y)}$\;
}
}
\Return $D'$\;
\end{algorithm}
The remaining question is how to compute $(X_i,Y_i)$ with minimum price (line 8) efficiently. We mention {\em densest subgraph in the bipartite graph} in the literature \cite{goldberg1984finding, khuller2009finding}, for which a poly-time algorithm based on max-flow has been proposed. 
\begin{definition}[Densest Subgraph in Bipartite Graph]
Given a bipartite graph $(X,Y,E)$ with $E: X \times Y \to \{0,1\}$, it asks to find $X' \subseteq X, Y'\subseteq Y$ to maximize $\frac{\sum_{x \in X', y \in Y'} E(x,y)}{|\rev{X'}| + |\rev{Y'}|}$.
\end{definition} 
Back to our problem, each value $b_i \in \dom(B)$ induces a bipartite graph with $X = \sigma_{B= b_i} R_1$, $Y = \sigma_{B =b_i} R_2$, and $E=\{(x,y)\mid x\in X, y\in Y, \pi_{A,C}(x\Join y)\notin \mathcal{C}\}$.
%$E = X \times Y - \mathcal{C}$.
This way, the densest subgraph in this bipartite graph corresponds to a subset of uncovered query results with minimum price, and vice versa. 

The approximation ratio of our greedy algorithm follows the standard analysis of weighted set cover~\cite{vazirani2001approximation}. We next focus on the time complexity. The while-loop proceeds in at most $O(|Q(D)|)$ iterations, since $|\mathcal{C}|$ increases by at least $1$ in every iteration. It takes $O(N)$ time to find the pair with minimum price, since there are $O(N)$ distinct values in $\dom(B)$. Moreover, it takes polynomial time to update $(X_i, Y_i)$ for each $b_i \in \dom(B)$. Overall, Algorithm~\ref{alg:greedy-matrix} runs in polynomial time in terms of $N$.

\subparagraph{Extensions.} Our algorithm for $\matrix$ can be extended to star CQs, and further to all CQs with only one non-output attribute. Similar property as Lemma~\ref{lem:decomposable} also holds. Our greedy algorithm needs a generalized primitive, noted as {\em densest subgraph in hypergraph}\footnote{Given a hypergraph $H = (V,E)$ for $E \subseteq 2^V$, it asks to find a subset of nodes $S \subseteq V$ such that the ratio $|\{e \in E: e\subseteq S\}|/|S|$ is maximized.}~\cite{hu2017maintaining} for finding the ``set'' with the smallest price. Following the similar analysis, we obtain:
\begin{theorem}
\label{lem:approximate-star}
For any CQ $Q$ with $|\attr(Q) - \head(Q)| =1$, $\ourprob$ is $O(\log N)$-approximable. 
\end{theorem}

\subsection{Line CQs}
\label{sec:line}
We turn to another commonly-used class of CQs noted as {\em line} CQs: \[Q_\textrm{line}(A_1, A_{m+1}):- R_1(A_1, A_2), R_2(A_2, A_3), \cdots, R_m(A_m,A_{m+1})\] with $m \ge 3$. We surprisingly find that \ourprob\ for line CQs is closely related to {\em directed Steiner forest} (DSF) problem~\cite{dodis1999design, chekuri2011set, feldman2012improved, iritETH2018} in the network design: Given an edge-weighed directed graph $G = (V, E)$ of $|V| = n$ and a set of $k$ demand pairs $\{(s_i, t_i) \in V \times V: i\in [k]\}$ and the goal is to find a subgraph $G'$ of $G$ with minimum weight such that there is a path in $G'$ from $s_i$ to $t_i$ for every $i \in [k]$. Observe that $\ourprob(Q_\textrm{line}, D)$ is a special case of DSF. This way, all existing algorithm proposed for DSF can be applied to \ourprob\ for $Q_\textrm{line}$. The best approximation ratio achieved is $O(\min\{k^{\frac{1}{2} + o(1)}, n^{0.5778}\})$~\cite{feldman2012improved, chekuri2011set, abboud2018reachability}. Combining the baseline in Section~\ref{sec:baseline} (with $\rho^* =2$ for line CQs) and existing algorithms for DSF, we obtain:

\begin{theorem}
\label{the:line-up}
For $Q_\textrm{line}$ and any database $D$ of input size $N$, there is a poly-time algorithm that can approximate $\ourprob(Q_\textrm{line}, D)$ within a factor of $O(\min\{|Q_\textrm{line}(D)|^{\frac{1}{2} + o(1)}\!, \dom^{0.5778}\!, N^{\frac{1}{2}}\})$, where $\dom$ is the number of values that participates in at least one full join result.
\end{theorem}
There is a large body of works investigating the lower bounds of DSF; and we refer interested readers to \cite{iritETH2018} for details. We mention a polynomial lower bound $\Omega(k^{1/4 - o(1)})$ for DSF, but it cannot be applied to \ourprob, since \ourprob\ is a special case with its complexity measured by the number of edges in the graph. Instead, we built a reduction from {\em label cover} to \ourprob\ for $Q_\textrm{line}$ directly (which is an adaption of reduction proposed in \cite{dodis1999design, chitnis2021parameterized}) and prove the following: 

\begin{theorem}
\label{the:line-lb}
No poly-time algorithm approximates $\ourprob$ for $Q_\textrm{line}$ within $\Omega(2^{(\log N)^{1-\epsilon}})$ factor for any constant $\epsilon > 0$, unless $\texttt{P} = \texttt{NP}$. \end{theorem}

\section{Related Work}
\subparagraph{Factorized database~\texorpdfstring{\cite{olteanu2016factorized}}{[35]}.} Factorized database studies a nested representations of query results that can
be exponentially more succinct than flat query result, which has the same goal as \ourprob. They built a tree-based representations by exploiting the distributivity of product over union and commutativity of product and union. This notion is quite different from \ourprob. They measure the regularity of factorizations by readability, the minimum over all its representations of the maximum number of occurrences of any tuple in that representation, while \ourprob\ measures the size of witness. 

\subparagraph{Data Synopses~\texorpdfstring{\cite{cormode2011synopses, phillips2017coresets}}{[15, 37]}.} In approximate query processing, people studied a lossy, compact synopsis of the data such that queries can be efficiently and approximately executed against the synopsis rather than the entire
dataset, such as random samples, sketches, histograms and wavelets. These data synopses differ in terms of what class of queries can be approximately answered, space usage, accuracy etc. In computational geometry, a corset is a small set of points that approximate the shape of a larger point set, such as for shape-fitting, density estimation, high-dimensional vectors or points, clustering, graphs, Fourier transforms etc. \ourprob\ can also be viewed as data synopsis, since it also selects a representative subset of tuples. But, it is more query-dependent, as different CQs over the same database (such as $Q_1(x):-R_1(x,y), R_2(y,z)$ and $Q_2(x,y,z):-R_1(x,y), R_2(y,z)$) can have dramatically different \ourprob.  Furthermore, all query results must be preserved by \ourprob, but this is never guaranteed in other data synopses. There is no space-accuracy tradeoff in \ourprob\ as well.  

\subparagraph{Related to Other Problems in Database Theory.} The \ourprob\ problem also outputs the smallest provenance that can explain why all queries results are correct. This notion of why-provenance has been extensively investigated in the literature~\cite{buneman2001and, green2007provenance, amsterdamer2011provenance}, but not from the perspective of minimizing the size of witness.  The \ourprob\ problem is also related to the resilience problem~\cite{FreireGIM15, freire2020new, qin2022resilience}, which intuitively finds the smallest number of tuples to remove so that the query answer turns into false. Our \ourprob\ problem essentially finds the maximum number of tuples to remove while the query answer does not change. We can observe clear connection here, but their solutions do not imply anything to each other. 

\section{Conclusion} 
In this paper, we study the data complexity of \ourprob\ problem for CQs without self-joins. There are several interesting problems left:
\begin{enumerate}
\item {\em Approximating \ourprob\ for CQs without head-domination property.} So far, the approximation of \ourprob\ is well-understood only on some specific class of CQs without head-domination property. For remaining CQs, both upper and lower bounds remain to be improved, which may lead to fundamental breakthrough for other related problems, such as DSF. 
\item {\em \ourprob\ for CQs with self-joins.} It becomes much more challenging when self-joins exists, as one tuple appears in multiple logical copies of input relation. \rev{Similar observation has been made for the related resilience problem~\cite{freire2020new, kimelfeld2012maximizing}}\item {\em Relaxing the number of query results witnessed.} It is possible to explore approximation on the number of query results that can be witnessed. Here, \ourprob\ is related to the {\em partial set cover} problem, for which many approximation algorithms~\cite{gandhi2004approximation} have been studied. 
\end{enumerate}

\bibliographystyle{plainurl}
\bibliography{p024-Hu}
\appendix
\section{Discussion on Our Improvement over Naive Solutions}

At last, let's see what can be benefited from our algorithmic results such as in Examples~\ref{ex:1}, \ref{ex:2} and \ref{ex:3}.
Below, we compare $\ourprob(Q,D)$ with $|D|$ and $|Q(D)|$.

For a CQ $Q$ with head-cluster property, let $Q_1$, $Q_2$, $\cdots$, $Q_k$ be the subqueries defined on relations partitioned by the output attributes.  For a CQ $Q$ with head-domination property, let $Q_1$, $Q_2$,$\cdots$, $Q_k$ be the subqueries defined by the connected components of $G^\exists_Q$. Consider an arbitrary database $D$. Let $D'_i$ be the projection of $Q(D)$ onto output attributes of $Q_i$. Let $D' = \sum_{i=1}^k D'_i$. We can prove for any $D$: (i) $|D'| \le |Q(D')| = |Q(D)|$; (ii) $\ourprob(Q,D) = |D'|$, if $Q$ has head-cluster property; (iii) $\texttt{ASWP}(Q,D) \le |\rel(Q)| \cdot |D'|$ if $Q$ has head-domination property. In general, $|Q(D)|$ can be as large as $\Theta\left(|D'|^\rho\right)$, where $\rho$ is fractional edge covering number of the residual CQ by removing all non-output attributes from $Q$, and $|D|$ can also be polynomially larger than $|D'|$, so $(\texttt{A})\ourprob(Q,D)$ could be polynomially smaller than both $|D|$ and $|Q(D)|$, while never be larger than them by a query-dependent constant.

For arbitrary CQ $Q$ with database $D$, let $D'$ be the projection of $Q(D)$ onto output attributes for each relation. We have proved that $\texttt{A} \ourprob(Q,D)$ is always smaller than $|D|$ and no larger than $|Q(D')|$ by a constant factor. The improvement would heavily depends on the specific database. As noted, in some applications (such as Example~\ref{ex:2} and \ref{ex:3}), a naive solution of size $|Q(D)|$ does not exist.

\section{Deletion Propagation}
\label{appendix:deletion-propagation}
In the literature, the classic deletion propagation (\DP) of CQs without self-joins show resemblance with \ourprob\ studied in this work. Here, we perform a comprehensive comparison between them with respect to problem formulation, complexity results and techniques. 

    \begin{definition}[Deletion Propagation with Side Effect~\cite{kimelfeld2012maximizing, kimelfeld2013multi}] Given a CQ $Q$, a database $D$ and a subset of query results $J \subseteq Q(D)$, the problem asks to find a subset of tuples $D' \subseteq D$ such that $J \cap Q(D-D') = \emptyset$, and there exists no subset $D'' \subseteq D$ with $J \cap Q(D-D'') = \emptyset$ and $|Q(D-D'')| > |Q(D-D')|$.
\end{definition}

\DP\ with side effects is different from \ourprob\ in terms of objective function, where \DP\ focuses on the number of remaining query results, and \ourprob\ focuses on the number of input tuples.

    \begin{definition}[Deletion Propagation with Source Side Effect~\cite{buneman2002propagation, cong2006annotation}]
        Given a CQ $Q$, a database $D$ and a subset of query results $J \subseteq Q(D)$, the problem asks to find a subset of tuples $D' \subseteq D$ such that all query results $J$ just disappear after removing all tuples in $D'$, i.e., $J \cap Q(D-D') = \emptyset$, while there exists no subset $D''\subseteq D$ with $J \cap Q(D-D'') = \emptyset$ and $|D''| < |D'|$.
    \end{definition}

    We note that if $J = \emptyset$, i.e., no query result is desired to be removed, the trivial solution $D'= \emptyset$ is optimal, and if $J = Q(D)$, \DP\ with source side effect just degenerates to the resilience problem (see Definition~\ref{def:resilience}). 

    \begin{definition}[Aggregated Deletion Propagation (\ADP)~\cite{hu2020aggregated}] Given a CQ $Q$, a database $D$ and a parameter $1 \le k \le |Q(D)|$, the problem asks to find a subset of tuples $D' \subseteq D$ such that at least $k$ query results disappear, i.e., $|Q(D-D')|\le |Q(D)|-k$, while there exists no subset $D''\subseteq D$ with  $|Q(D-D'')|\le |Q(D)|-k$ and $|D''| < |D'|$.
    \end{definition}

    Similarly, we note that if $k = 0$, i.e., no query result is desired to be removed, the trivial solution $D'= \emptyset$ is optimal, and if $k = |Q(D)|$, \ADP\ degenerates to the resilience problem. We next focus on the difference between resilience and \ourprob:
    
    \begin{definition}[Resilience~\cite{FreireGIM15, freire2020new}]
    \label{def:resilience}
    Given a Boolean CQ $Q$ and a database $D$ with $Q(D)$ is true, the problem asks to find a subset of tuples $D' \subseteq D$ such that $Q(D-D')$ is false, while there exists no subset $D''\subseteq D$ with $Q(D-D'')$ as false and $|D''| < |D'|$.   
    \end{definition}

    These two problems have completely different constraints on $D'$, such that $Q(D') = Q(D)$ is always required by \ourprob, but $Q(D - D') = \emptyset$ is required by the resilience problem. Note that $Q(D - D') \neq Q(D) - Q(D')$! Hence, $Q(D - D') = \emptyset$ does not imply anything for $Q(D')$, so there is no connection between solutions to the resilience problem and solutions to \ourprob.

    As head domination property also precisely captures the hardness of deletion propagation with side effect, we next focus on the difference between our paper and \cite{kimelfeld2012maximizing}:
    \begin{itemize}
        \item Poly-time solvability (assume $\texttt{P} \neq \texttt{NP}$):
        \begin{itemize}
            \item \ourprob\ is poly-time solvable if and only if $Q$ has head-cluster property;
            \item \DP\ is poly-time solvable if and only if $Q$ has head-domination property;
        \end{itemize}
        \item Approximability:
        \begin{itemize}
            \item \ourprob\ is $N^{1-1/\rho}$-approximable for any $Q$; 
            \item \DP\ is $1/\min\{|\attr(Q)|, |\rel(Q)|\}$-approximable for any $Q$;
        \end{itemize}
        \item Inapproximability for CQ without head-domination property:
        \begin{itemize}
            \item \ourprob\ is NP-hard to approximate within a factor of $(1-o(1))\cdot \log N$;
            \item \ourprob\ is NP-hard to approximate for line CQs within a factor of $O\left(2^{(\log N)^{1-\epsilon}}\right)$ for any constant $\epsilon > 0$;
            \item \DP\ is NP-hard to approximate within a factor of $\alpha$, for some constant $\alpha \in (0,1)$, which may vary over different CQs;
        \end{itemize}
        \item Greedy Algorithms:
        \begin{itemize}
        \item \ourprob\ is $O(\log N)$-approximable for star CQs.
        \item \ourprob\ is $O(\min\{|Q_\textrm{line}(D)|^{\frac{1}{2} + o(1)}, \dom^{0.5778}, N^{\frac{1}{2}}\})$-approximable for line CQs (see Theorem~\ref{the:line-up}).
        \item \DP\ is $\frac{1}{2}$-approximable for star CQs.
        \end{itemize}
    \end{itemize}

     We are not aware of any similar techniques used for algorithms and lower bounds between this paper and \cite{kimelfeld2012maximizing}.

\section{Missing Materials in Section~\ref{sec:exact}}
\label{appendix:SWP-easy}

    \rev{
    
    \begin{algorithm}[t]
	\caption{{\sc EasySWP}$(Q,D)$}
	\label{alg:easy-SWP-rewrite}
	$D' \gets \emptyset$\;
	$\{E_1, \cdots, E_k\} \gets $ a partition of $\rel(Q)$ by output attributes\;
        ${\bf A}_1, {\bf A}_2, \cdots, {\bf A}_k \gets $ output attributes of $E_1, E_2, \cdots, E_k$\;
	\ForEach{$i \in [k]$}{
                Define $Q_i({\bf A}_i):- \{R_j(\allattr_{j}):j \in E_i\}$\;
            \ForEach{$t' \in \pi_{{\bf A}_i} Q(D)$}{
	      $D' \gets D' \biguplus \ourprob(Q_i,  \{R_j: R_j \in E_i\}, t')$\;
	    }
	}
	  \Return $D'$\;
	\end{algorithm}
 }

    \subsection{Faster Algorithm of Computing \ourprob\ for Triangle CQ} Let's take the triangle full CQ $Q_\triangle(A,B,C) = R_1(A,B) \Join R_2(B,C) \Join R_3(A,C)$ as example. For an arbitrary database $D$, $\ourprob(Q_\triangle,D)$ can be computed in $O\left(N^{\frac{2\omega}{\omega+1}}\right)$ time, where $\omega$ is the exponent of fast matrix multiplication algorithm.  The algorithm is adapted from triangle detection. For each input relation, we proceed with the following procedure. Let's take $R_1$ as example. 
    
    This algorithm uses a parameter $\Delta$, whose value will be determined later. A value $a \in \dom(A)$ is {\em heavy} if it appears in more than $\Delta$ tuples in $R_1(A,B)$ or $R_3(A,C)$, and {\em light} otherwise. A value $b \in \dom(B)$ is {\em heavy} if it appears in more than $\Delta$ tuples in $R_1(A,B)$ or $R_2(B,C)$, and {\em light} otherwise. A value $c \in \dom(C)$ is {\em heavy} if it appears in more than $\Delta$ tuples in $R_2(B,C)$ or $R_3(A, C)$, and {\em light} otherwise.  We take the union of results of the following queries:
    \begin{itemize}
        \item $\pi_{A,B} \left(R_1(A,B^{\textrm{light}}) \Join R_2(B^{\textrm{light}},C) \ltimes R_3(A,C)\right)$: We first materialize $R_1(A,B^{\textrm{light}}) \Join R_2(B,C)$, then compute the semi-join between the intermediate join results and $R_3$, and the projection of the semi-join results onto attributes $A,B$. It takes $O(N \cdot \Delta)$ time.
        \item $\pi_{A,B} \left(R_1(A^{\textrm{light}}, B) \Join R_3(A^{\textrm{light}},C) \ltimes R_2(B,C)\right)$: We apply similar procedure as above.  It also takes $O(N \cdot \Delta)$ time.
        \item $R_1(A,B) \ltimes \left(R_2(B, C^{\textrm{light}})\Join R_3(A,C^{\textrm{light}})\right)$:  We materialize $R_2(B,\\ C^{\textrm{light}})\Join R_3(A,C^{\textrm{light}})$, then compute the semi-join between $R_1$ and the intermediate join results. It takes $O(N \cdot \Delta)$ time.
        \item $R_1(A,B) \Join \left(R_2(B^{\textrm{heavy}}, C^{\textrm{heavy}})\Join R_3(A^{\textrm{heavy}},C^{\textrm{heavy}})\right)$:  We materialize $R_2(B^{\textrm{heavy}}, C^{\textrm{light}}) \Join R_3(A^{\textrm{heavy}},C^{\textrm{light}})$ using fast matrix multiplication, then compute the semi-join between $R_1$ and the intermediate join results. It also takes $O\left((\frac{N}{\Delta})^{\omega}\right)$ time.
    \end{itemize}
    %Each query can be computed in $O\left(N \cdot \Delta + (\frac{N}{\Delta})^{\omega}\right)$ time. 
    By setting $\Delta = N^{\frac{\omega-1}{\omega+1}}$, we obtain the overall run-time as $O(N^{\frac{2\omega}{\omega+1}})$.
    
\section{Missing Proofs in Section~\ref{sec:approximate}}
\label{appendix:approximate}

\begin{proof}[Missing Details in the Proof of Lemma~\ref{lem:matrix-integral}]
Here, we show that $(\mathcal{U}, \mathcal{S})$ has a cover of size $\le k$ if and only if the $\ourprob(\matrix, D)$ has an integral solution of size $\le |\mathcal{U}| + k \cdot |V| = n(k+1)$.

 {\em Direction only-if.} Suppose we are given a cover $\mathcal{S}'$ of size $k$ to $(\mathcal{U}, \mathcal{S})$. We construct an integral solution $D'$ to $\ourprob(\matrix, D)$ as follows. Let $B' \subseteq \dom(B)$ be the set of values that corresponding to $\mathcal{S}'$. For every $b_S \in B'$, i.e., $S \in \mathcal{S}'$, we add tuple $(b_S, c_u)$ to $D'$ for every $u \in \mathcal{U}$. For every $a_u \in \dom(A)$, we choose an arbitrary value $b_S \in B'$ such that $u \in S$, and add tuple $(a_u, b_S)$ to $D'$. This is always feasible since $\mathcal{S}'$ is a valid set cover. It can be easily checked that $n$ tuples from $R_1$ and $k \cdot n$ tuples from $R_2$ are added to $D'$.
       
 {\em Direction if.} Suppose we are given a integral solution $D'$ of size $k'$ to $\ourprob(\matrix, D)$. Let $B'$ be the subset of values whose incident tuples in $R_2$ are included by $D'$. We argue that all subsets corresponding to $B'$ forms a valid cover of size $\frac{k'}{n-1}$. By definition of integral solution, $|B'| = \frac{k'}{n} -1$. Moreover, for every value $a \in \dom(A)$, at least one edge $(a,b)$ is included by $D'$ for some $b \in B'$, hence $B'$ must be a valid set cover. 
\end{proof}

\begin{definition}[Cycle]
\label{def:cycle}
    In a CQ $Q$, a sequence of attributes $P = \langle A_1, A_2, \cdots, A_k\rangle$ is a cycle if
    \begin{itemize}
        \item for every $i \in [k]$, there exists a relation $R_j \in \rel(Q)$ such that $A_i, A_{(i+1) \mod k} \in \attr(R_j)$;
        \item for every $R_j \in \rel(Q)$, either $\attr(R_j) \cap P = \emptyset$, or $|\attr(R_j) \cap P |=1$ or $\attr(R_j) \cap P = \{A_i, A_{(i+1)\mod k}\}$ for some $i \in [k]$.
    \end{itemize}
\end{definition}

\begin{definition}[Non-Conformal Clique]
\label{def:clique}
    In a CQ $Q$, a non-conformal clique is a subset of attributes $P \subseteq \attr(Q)$, such that 
    \begin{itemize}
        \item for every pair of attributes $A_i,A_j$, there exists a relation $R_k \in \rel(Q)$ with $A_i,A_j \in \attr(R_k)$;
        \item there is no relation $R_\ell \in \rel(Q)$ such that $P \subseteq \attr(R_\ell)$.
    \end{itemize}
\end{definition}

\begin{lemma}[\cite{brault2016hypergraph}]
\label{lem:cyclic}
    Every cyclic CQ contains either a cycle or a non-conformal clique. 
\end{lemma}

\begin{proof}[Proof of Lemma~\ref{lem:connection}]
    We already prove this lemma for acyclic CQs, such that if there exists no free sequence, $Q$ has head-domination property. Below, we focus on cyclic CQs. From Lemma~\ref{lem:cyclic}, $Q$ must contain a cycle or a non-formal clique. 
    
    Consider an arbitrary cycle or clique $P$. As shown in Lemma~\ref{lem:cycle-clique-head}, if $P \cap \head(Q) \neq \emptyset$ and $P - \head(Q) \neq \emptyset$, there must exist a relation $R_k \in \rel(Q)$ such that removing $\head(R_k)$ turns $Q$ into a disconnected CQ, where relations in $\{R_i: \attr(R_i) \cap P - \head(Q) \neq \emptyset\}$ is a connected subquery. Then, it is safe to remove all relations in $\{R_i: \attr(R_i) \cap P - \head(Q) \neq \emptyset\}$ from $Q$, and prove that the residual query (where $P$ disappears) has head-domination property.    
    
    Applying this removal procedure iteratively, we are left with cycles and non-conformal cliques that contain only output attributes, or only non-output attributes. Lemma~\ref{lem:cycle-clique} proves that $Q$ in this case has head-domination property, thus completing the proof.
\end{proof}

\begin{lemma}
\label{lem:cycle-clique}
    In a cyclic CQ $Q$ without free sequences and nested cliques in its renamed query, if every cycle or non-conformal clique contains only output attributes, or only non-output attributes, $Q$ has head-domination property.
\end{lemma}

\begin{proof}[Proof of Lemma~\ref{lem:cycle-clique}]
We apply the following procedure to $Q$: 
if there exists a cycle or non-conformal clique $P$, we add a {\em virtual} relation $R_p$ with $\attr(R_p) = P$ to $\rel(Q)$. We apply this procedure iteratively until no more cycle or non-conformal clique can be found. Let $Q'$ be the resulted query.  

We first prove that for each virtual relation $R_p$ added, either $\attr(R_p) \subseteq \head(Q)$ or $\attr(R_p) \subseteq \attr(Q) - \head(Q)$.
Adding virtual relation does not increase more cycles, but some possible non-conformal cliques. We distinguish the following three cases: 
\begin{itemize}
    \item If $R_p$ is added for a cycle $P$, then $P$ must exist in $Q$ and therefore $\attr(R_p) \subseteq \head(Q)$ or $\attr(R_p) \subseteq \attr(Q) - \head(Q)$;
    \item If $R_p$ is added for a non-conformal clique $P$, and $P$ exists in $Q$, then $\attr(R_p) \subseteq \head(Q)$ or $\attr(R_p) \subseteq \attr(Q) - \head(Q)$;
    \item Otherwise, $R_p$ is added for a non-conformal clique $P$ but $P$ does not exist in $Q$. It must be the case that after $R_{p'}$ is added for a cycle $P'$, a non-conformal clique $P$ forms with $P'\subsetneq P$. If $P'\subseteq \head(Q)$, then $P \subseteq \head(Q)$; otherwise, $P$ is a nested clique in the renamed query of $Q$. If $P'\subseteq \attr(Q) - \head(Q)$, then $P \subseteq \attr(Q) - \head(Q)$; otherwise, any pair of non-consecutive attributes in $P'$ will form a free sequence together with an arbitrary attribute in $P \cap \head(Q)$, contradicting the fact that $Q$ does not contain a free sequence. In either case, $R_p$ is added as a virtual relation with $\attr(R_p) \subseteq \head(Q)$ or $\attr(R_p) \subseteq \attr(Q) - \head(Q)$. 
 \end{itemize}

 We next point out some critical observations on $Q'$: (1) $\head(Q) = \head(Q')$; (2) $Q'$ is acyclic, as it does not contain any cycle or non-conformal clique; and (3) $Q'$ also does not contain any free sequence as $Q$. Together, $Q'$ is free-connex. It is feasible to find a free-connex join tree, and identify connected components $E'_1, E'_2, \cdots, E'_k$ of $G^\exists_{Q'}$ with dominants $\dot{R}_1, \dot{R}_2, \cdots, \dot{R}_k \in \rel(Q')$. %5such that the following properties are satisfied for every $i \in [k]$:
 %\begin{itemize} 
 %   \item for any relation $R_j \in E'_i$, $\head(R_j) \subseteq \head(\dot{R}_i)$;
 %   \item removing attributes in $\head(\dot{R}_i)$ turns $Q'$ into a disconnected CQ, such that $E'_i \cup \{\dot{R}_i\}$ form a connected subquery. 
%\end{itemize}
Moreover, as $Q'$ is connected, $\head(\dot{R}_i) \neq \emptyset$ holds for every $i\in [k]$. We next show that virtual relations are not dominants. By contradiction, we assume $R_p = \dot{R}_1$ for some virtual relation $R_p$. If $\attr(R_p) - \head(Q) = \emptyset$,  $E'_1 \cup \{R_p\}$ form a connected subquery, so $Q'$ is disconnected, leading to a contradiction. If $\attr(R_p) \subseteq \head(Q)$, we further distinguish two more cases. If $R_p$ is added for a non-conformal clique, then $E'_1$ form a nested-clique in the rename query of $Q$; and if $R_p$ is added for a cycle, then any non-consecutive attributes of $P$ together with non-output attributes in relations of $E'_1$ form free-sequence. Hence, virtual relations cannot be dominants, and $\dot{R}_1, \dot{R}_2, \cdots, \dot{R}_k \in \rel(Q)$. 

Let $E_i\subseteq E'_i$ be the set of non-virtual relations in $E'_i$. It can be easily checked that $E_1,E_2,\cdots, E_k$ are also the connected components of $G^\exists_Q$ with dominants $\dot{R}_1, \dot{R}_2, \cdots, \dot{R}_k$. Hence, $Q$ has head-domination property.
\end{proof}

\begin{algorithm}[t]
	\caption{{\sc GreedySWP}$(Q,D)$}	\label{alg:baseline}
	$D' \gets \emptyset$\;
	\lForEach{$t \in Q(D)$}{$D' \gets D' \cup \ourprob(Q,D, t)$}
	\Return $D'$\;
\end{algorithm}
\begin{lemma}
\label{lem:cycle-clique-head}
    In a cyclic CQ $Q$ without free sequences and nested cliques in its renamed query, for any cycle or non-conformal clique $P$, if $P \cap \head(Q) \neq \emptyset$ and $P - \head(Q) \neq \emptyset$, there must exist a relation $R_k \in \rel(Q)$ such that removing $\head(R_k)$ turns $Q$ into a disconnected CQ, where $\{R_i: \attr(R_i) \cap P - \head(Q) \neq \emptyset\}$ is a connected subquery.
\end{lemma}
\begin{proof}
    Consider an arbitrary non-output attribute $A \in \attr(Q) - \head(Q)$. Let $V_A\subseteq \attr(Q)$ be the set of attributes renamed to attribute $A$ in this procedure. Let $N_A = \{B \in \attr(Q) - V_A: \exists B' \in V_A, R_i \in \attr(Q), s.t. B,B' \in \attr(R_i)\}$ be the set of attributes that appear together with some attribute of $V_A$ in some relation.  Note that $N_A \subseteq \head(A)$; otherwise, any attribute in $N_A - \head(A)$ will be renamed as $A$, coming to a contradiction. 
    Moreover, for every pair of attributes $B,B' \in N_A$, there must exist a relation $R_i$ such that $B, B' \in \attr(R_i)$; otherwise, a free sequence is identified, coming to a contradiction. Thus, $N_A$ is a clique. In addition, there must exist one relation $R_k$ such that $N_A \subseteq \attr(R_k)$; otherwise, a nested clique $N_A \cup \{A\}$ is identified in the renamed query, coming to a contradiction. In this way, removing all attributes in $V_A \subseteq \head(R_k)$ turns $Q$ into a disconnected CQ, where $\{R_i \in \rel(Q): \head(R_i) \subseteq V_A\}$ is a connected subquery. 
    
    Let's come to cycle $P$. For simplicity, denote $P = \langle A_1,A_2,\cdots, A_k\rangle$. As there is no free sequence, $P \cap \head(Q) = \{A_i, A_{(i+1) \mod k}\}$ for some $i \in [k]$. Then, we can identify an arbitrary output attribute in $P - \head(Q)$ as $A$, and therefore
    \begin{itemize}
        \item $P - \head(Q) \subseteq V_A$;
        \item for every relation $R_j \in \rel(Q)$ with $\attr(R_j) \cap P -\head(Q) \neq \emptyset$, $\head(R_j) \subseteq V_A$. 
    \end{itemize}
    Hence, removing $\head(R_k)$ turns $Q$ into a disconnected CQ, where $\{R_i: \attr(R_i) \cap P - \head(Q) \neq \emptyset\}$ is a connected subquery. 

     We next turn to non-conformal clique $P$. We can identify an arbitrary output attribute in $P - \head(Q)$ as $A$, and therefore
    \begin{itemize}
        \item $P - \head(Q) \subseteq V_A$;
        \item for every relation $R_j \in \rel(Q)$ with $\attr(R_j) \cap P -\head(Q) \neq \emptyset$, $\head(R_j) \subseteq V_A$. 
    \end{itemize}
    Hence, removing $\head(R_k)$ turns $Q$ into a disconnected CQ, where $\{R_i: \attr(R_i) \cap P - \head(Q) \neq \emptyset\}$ is a connected subquery.  
\end{proof}

\section{Missing Proofs in Section~\ref{sec:approximate}}
\label{appendix:approximation-algorithm}

\begin{proof}[Proof of Theorem~\ref{the:baseline}]
    Let $D'$ be the solution returned by Algorithm~\ref{alg:baseline}. For each query result $t \in Q(D)$, we add at most $|\rel(Q)|$ tuples to $D'$, so $|D'|\leq |Q(D)| \cdot |\rel(Q)|$. Meanwhile, $|D'|\leq N \cdot |\rel(Q)|$. Let $D^*$ be the solution to \ourprob$(Q,D)$. Note that at most $|D^*|^{\rho}$ query results can be reproduced from $D^*$. Hence, we must have $|D^*|^{\rho} \ge |Q(D)|$, i.e., $|D^*| \ge |Q(D)|^{1/\rho}$. Below, we show that $|D'| \le N^{1-1/\rho} \cdot |D^*|$ with the following facts:
    \begin{itemize}
        \item If $|Q(D)|\leq N$, then the approximation ratio (skipping $|\rel(Q)|$) is $\frac{|D'|}{|D^*|} \le \frac{|Q(D)|}{|Q(D)|^{1/\rho}}=|Q(D)|^{1-1/\rho}\leq N^{1-1/\rho}$. 
        \item If $N<Q(D)$, then the approximation ratio (skipping $|\rel(Q)|$) is at most $\frac{|D'|}{|D^*|}  \le \frac{N}{|Q(D)|^{1/\rho}}\leq \frac{N}{N^{1/\rho}}=N^{1-1/\rho}$.
    \end{itemize}
    As $Q(D)$ can be computed in polynominal time, $|Q(D)|$ is polynomially large in terms of $|D|$, and \ourprob\ for a single tuple can be computed in polynominal time, the baseline runs in polynominal time. 
\end{proof}

\begin{proof}[Proof of Theorem~\ref{the:line-lb}]
We focus on line-3 query $Q_\textrm{line3}(A_1, A_4):-R_1(A_1, A_2), R_2(A_2, A_3), \\R_3(A_3,A_4)$, which can be generalized to line-$m$ query $Q_\textrm{line}(A_1, A_m):-R_1(A_1, A_2), R_2(A_2, A_3),\\ \cdots, R_m(A_m,A_{m+1})$, by putting $A_3,A_4,\cdots, A_{m-1}$ as identical copies of $A_m$.
  \begin{figure}[t]
        \centering
    \includegraphics[scale=0.9]{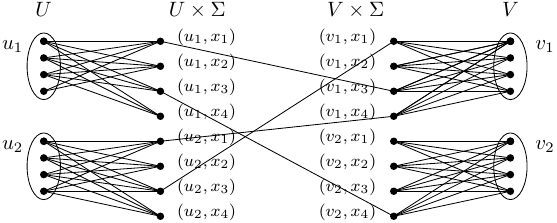}
        \caption{An example of hard instance constructed for $Q_\textrm{line3}$.}
        \label{fig:line3}
    \end{figure}
We introduce the {\em Label Cover Problem} as below:
\begin{definition}[Label Cover]
    Given a complete bipartite graph $(U,V,E)$ with $|U| = |V| =n$, a finite alphabet $\Sigma$ with $\Sigma > n$, constraints $C_{u,v}$ for every edge $e = (u,v)$, the goal is find assignments to the vertices $M_U: U \to 2^\Sigma$ and $M_V: V \to 2^\Sigma$ that every edge is satisfied, i.e., there exists $x \in M_U(u), y \in M_V(v)$ such that $(x,y) \in C_{u,v}$, and $\sum_{u \in U} |M_U(u)| + \sum_{v \in V}|M_V(v)|$ is minimized. The problem size $m$ is defined as $n \times |\Sigma|$.
\end{definition}

\begin{lemma}[\cite{dinur2014analytical}]
\label{lem:label-cover}
    There is no poly-time algorithm that can approximate the label cover instances of size $m$ within $O\left(2^{(\log m)^{1-\epsilon}}\right)$ factor for any $\epsilon > 0$.
\end{lemma}

    We set $\dom(A) = U \times [n^2]$, $\dom(B) = U \times \Sigma$, $\dom(C) = V \times \Sigma$ and $\dom(D) = V \times [n^2]$.
    The database $D$ includes:
    \begin{itemize}
        \item $R_1 =\{(\langle u,i \rangle, \langle u,x \rangle): u\in V, i \in [n^2], x \in \Sigma\}$;
        \item $R_2 = \{(\langle u,x \rangle, \langle v,y \rangle): u,v\in V, x,y\in \Sigma, (x,y) \in C_{uv}\}$;
        \item $R_3 = \{(\langle v,y \rangle, \langle v,i \rangle ): v\in V, y \in \Sigma, i \in [n^2]\}$;
    \end{itemize}
    In this instance, we have $N = O(n^2 \cdot |\Sigma|^2 + n^3 \cdot |\Sigma|) = O(m^2)$. It can be easily checked that $\line(D) = U \times [n^2] \times V \times [n^2]$ with $n^6$ query results.
    For any witness $D'$ for $\ourprob(\line, D)$, let $D'_1, D'_2, D'_3$ be the set of tuples chosen from $R_1, R_2, R_3$ separately. 

    For a sub-database $D' \subseteq D$, let $R'_1, R'_3$ are the corresponding subset of tuples from $R_1, R_3$ in $D'$.
    A solution $D'$ to $\ourprob(\line, D)$ is integral if for any $(u,x) \in U \times \Sigma$, either $\sigma_{B = (u,x)} R'_1 = \emptyset$ or $\sigma_{B = (u,x)} R'_1 = \{\langle u,i\rangle, \langle u,x\rangle: i \in [n^2]\}$, and for any $(v,y) \in V \times \Sigma$, either $\sigma_{C = (v,y)} R'_3 = \emptyset$ or $\sigma_{C = (v,y)} R'_3 = \{\langle v,i\rangle, \langle v,y\rangle: i \in [n^2]\}$.
    
    \begin{lemma}
        Every non-integral solution to $\ourprob(\line, D)$ can be transformed into an integral solution to $\ourprob(\line, D)$.
    \end{lemma}
    
    \begin{proof}
        Consider an arbitrary non-integral solution $D'$. W.o.l.g., assume there exists some $(u,x) \in U \times X$, such that $\sigma_{B = (u,x)} R'_1 \neq \emptyset$ and $ \sigma_{B = (u,x)} R'_1 \neq \{\langle u,i\rangle, \langle u,x\rangle: i \in [n]\}$. There must exist some pair of $(i,j)$ such that $\langle u,i\rangle, \langle u,x\rangle \in D'_1$ but $\langle u,j\rangle, \langle u,x\rangle \notin D'_1$. Let $X_i =\{x\in X: \langle u,i\rangle, \langle u,x\rangle \} \in D'_1$ and $X_j  = \{x\in X: \langle u,i\rangle, \langle u,x\rangle \} \in D'_1$. W.l.o.g., assume $|X_i| \le |X_j|$. As $D'$ is a solution to $\ourprob(\line, D)$, it is feasible to remove $\langle u,j\rangle, \langle u,x\rangle \in D'_1$ for $x \in X_j$ from $D'$ and add $\langle u,j\rangle, \langle u,x\rangle \in D'_1$ for $x \in X_i$ to $D'$, which yields another solution to $\ourprob(\line, D)$ without increasing the number of tuples. After applying this step iteratively, we can obtain a solution such that every $(u,i)$ is incident to the same values in $\dom(B)$ over $i \in [n]$, i.e., an integral solution. 
    \end{proof}

    From now on, it suffices to consider the integral solution to \ourprob$(\line, D)$.  Moreover, we can assume $|D'_2| = n^2$ by picking an arbitrary edge $(\langle u,x\rangle, \langle v,y\rangle) \in R_2$ with $(x,y) \in C_{uv}$ for $(u,v) \in U \times V$. Observe that there is a solution to label cover of cost at most $c$ if and only if there is an integral solution to $\ourprob$ for $\line$ of cost at most $n^2 c + n^2 = n^2(c+1)$. 
    \begin{itemize}
        \item {\em Direction Only-If:} Let $S = (M_U, M_V)$ be the solution to label cover. We define an integral solution $D_S$ to $\ourprob(\line, D)$ based $S$. For each vertex $u \in U$, we add the set of tuples $(\langle u, * \rangle, x) \in R_1$ to $D_S$ if $x\in M_U(u)$. Similarly, for each vertex $v \in V$, we add the set of tuples $(\langle v, * \rangle, y) \in R_3$ to $D_S$ if $y \in M_V(v)$. Together with the $n^2$ tuples chosen from $R_2$, we have obtained a witness $D_S$ of size $c \cdot n^2 + n^2= (c+1)n^2$.
        \item {\em Direction If:} 
        Let $D_S$ be an integral witness to  $\ourprob(\line, D)$ of cost $c'$. We construct a solution $(M_U, M_V)$ to label cover as follows. If $(\langle u, * \rangle, x) \in R_1$ is included by $D_S$, we add $x$ to $M_U(u)$. Similarly, if $(\langle v, * \rangle, y) \in R_3$ is included by $D_S$, then we add $y$ to $M_V(v)$. It can be easily checked that we have obtained a solution to label cover of size $\frac{c'-1}{n^2}$. 
    \end{itemize}
    
    If there is a poly-time algorithm that can approximate $\ourprob$ for $\line$ within $2^{(\log \sqrt{N})^{1-\epsilon}}$ factor, then there is a poly-time algorithm that can approximate label cover within $2^{(\log \sqrt{N})^{1-\epsilon}} = 2^{(\log m)^{1-\epsilon}}$ factor, coming to a contradiction of Lemma~\ref{lem:label-cover}. 
    %Observe that there is a solution to directed steiner network of cost $c$ if and only if there is a solution to directed steiner network of cost $n^2(2c+1)$. Therefore, 
    %\begin{itemize}
    %    \item If $\textrm{val}(G) =1$, then there is an integral witness to $\ourprob(\line, D)$ of cost $3n^2$. 
    %    \item If $\textrm{val}(G) < \gamma$, then there is an integral witness to $\ourprob(\line, D)$ of cost more than $n^2 (2\sqrt{\frac{2}{\gamma}} + 1) \ge 3n^2 \sqrt{\frac{2}{\gamma}}$. 
    %\end{itemize}
    %More specifically, if $\textrm{val}(G) < 2^{(\log n)^{\frac{1}{2} + \rho}}/n$, then the optimal integral witness to $\ourprob(\line, D)$ has cost at least $3 n^2 \sqrt{\frac{2n}{2^{(\log n)^{\frac{1}{2} + \rho}}}}$. If the approximation ratio of $\ourprob(\line, D)$ is smaller than 
    %\[\frac{3 n^2 \sqrt{\frac{2n}{2^{(\log n)^{\frac{1}{2} + \rho}}}}}{3n^2} =  \sqrt{\frac{2n}{2^{(\log n)^{\frac{1}{2} + \rho}}}} = O\left(\frac{N^{1/8}}{2^{(\log n)^{\frac{1}{2} + \rho'}}}\right),\]
    %then it is possible distinguish these two cases of $G$, which contradicts Conjecture~\ref{con:2-CSP}.

    \end{proof}
\end{document}